\documentclass[fleqn,usenatbib]{mnras}

\usepackage{newtxtext,newtxmath}
\usepackage[T1]{fontenc}

\usepackage{booktabs}
\usepackage{graphicx}
\usepackage{multirow}
\usepackage{ulem}

\hypersetup{
  colorlinks=true,
  citecolor=blue,
  linkcolor=blue,
  urlcolor=blue
}

\newcommand{\e}{\mathrm{e}}
\newcommand{\dd}{\mathrm{d}}

\DeclareMathOperator{\Cov}{Cov}
\DeclareMathOperator{\diag}{diag}

\newcommand{\bK}{\boldsymbol{K}}
\newcommand{\by}{\boldsymbol{y}}
\newcommand{\bX}{\boldsymbol{X}}

\title[Intrinsic versus observation induced spin nonstationarity in pulsar timing]{Intrinsic versus observation induced spin nonstationarity in pulsar timing}

\author[R. C. Bernardo]{
Reginald Christian Bernardo,\thanks{E-mail: reginald.christian.bernardo@aei.mpg.de}\\
Max Planck Institute for Gravitational Physics (Albert Einstein Institute), \\ Leibniz Universit\"{a}t Hannover, Callinstrasse 38, D-30167 Hannover, Germany
}

\date{Accepted XXX. Received YYY; in original form ZZZ}
\pubyear{\the\year{}}

\begin{document}
\label{firstpage}
\pagerange{\pageref{firstpage}--\pageref{lastpage}}
\maketitle

\begin{abstract}
Whether spin noise of a neutron star is stationary is, in two-component models, a question of whether stellar angular momentum is conserved internally or exchanged externally through independent stochastic torques on the crust and superfluid. The question is posed in the two components' angular velocities, but radio pulsar timing observes only the crust, through its rotational phase after a deterministic timing model is fitted. We develop a Gaussian process method for testing spin noise stationarity directly with such phase data. Starting from the exact analytical phase means and covariances of a singular, nonstationary and a minimal, stationary two-component model, we show that the crust phase covariances are semiseparable and construct a likelihood that scales linearly with the number of observations, with the timing model marginalized analytically. We analyze simulated phase data from both models using full-state and crust-only observations, and after differentiation into local spin frequencies. Integration, timing model removal and differentiation neither create nor destroy the distinction between stationary and nonstationary spin noise, but determine how much survives in a finite, noisy data set. With both components observed, the models are distinguished decisively and the generating parameters are recovered. With the crust alone, both models describe the observed phase equally well over much of parameter space, and the Bayes factors are of order unity and prior-limited. The minimal model nonetheless ties the hidden superfluid phase to the observed crust phase through a single factor set by the superfluid inertia fraction, so an independent estimate of this fraction, as from glitches, turns it into a falsifiable prediction that a continuous gravitational wave observation of the interior could test.
\end{abstract}


\section{Introduction}
\label{sec:introduction}

Radio pulsars \citep{Hewish:1968bj,Pilkington:1968bk} are exceptionally stable rotators, and the precision with which their pulses can be timed has made them laboratories for fundamental physics \citep{Lorimer:2001vd,Manchester:2015mda,Verbiest:2024nid}. Timing of pulsars in relativistic binaries provides strong-field tests of general relativity \citep{Hulse:1974eb,Kramer:2021jcw}, and the timing of arrays of millisecond pulsars has recently yielded support for a nanohertz gravitational wave background \citep{NANOGrav:2023gor,EPTA:2023fyk,Reardon:2023gzh,Xu:2023wog, Miles:2024seg, 2025arXiv251208666Y}. This precision is nonetheless bounded by timing noise, a slow, red wandering of the rotational phase relative to a deterministic spin-down model that was recognized early in the Crab pulsar \citep{1975ApJS...29..453G} and that is present in most pulsars with long timing baselines \citep{2010MNRAS.402.1027H}. Timing noise limits the low-frequency sensitivity of pulsar timing experiments \citep{2010ApJ...725.1607S, 2015MNRAS.449.3293L}, but it is also a signal in its own right, since it carries information on the rotational dynamics of the neutron star \citep{Melatos:2013rca, Goncharov:2019mei}.

A physical picture of timing noise is provided by two-component models of the neutron star, in which a crust that is electromagnetically coupled to the observed pulses exchanges angular momentum with an interior superfluid through mutual friction, and in which stochastic torques act on one or both components \citep{Meyers:2021myb,Meyers:2021slh,Antonelli:2022gqw}. The same two-component picture underlies the standard interpretation of glitches \citep{1969Natur.224..872B,1975ApJS...29..453G,2012PhRvL.109x1103A, 2013PhRvL.110a1101C, Haskell:2015jra}.
The coupling time scale has recently been constrained from timing noise, first for PSR J1359-6038 \citep{ONeill:2024uiw} and subsequently for a sample of UTMOST pulsars \citep{2019MNRAS.484.3691J, 2020MNRAS.494..228L, 2025MNRAS.541.1792D, Dong:2025nho}.
Within this picture a physical question arises that the present work is concerned with. If the stochastic torques on the two components are independent, the total angular momentum of the star performs a random walk and the spin noise is nonstationary \citep{Meyers:2021myb}. If instead the stochastic torques are internal, equal and opposite between the components, the total angular momentum is conserved and the spin noise of each component is stationary \citep{Antonelli:2022gqw}. The stationarity of pulsar spin noise is therefore a statement about whether angular momentum is conserved inside the neutron star, and it can in principle be tested by comparing two models, referred to below as singular and minimal, that realize the two alternatives.

This work follows up \cite{Bernardo:2026opv}, which obtained exact time-domain solutions for the singular two-component spin-noise model \citep{Meyers:2021myb} and established the stochastic framework used here.
In a different work, we take up the physical question directly in the variables in which it is posed, the angular velocities or spin frequencies of the crust and the superfluid, and develops scalable Gaussian process methods to test spin noise stationarity with angular velocity residuals.
The present work extends the analytical treatment to the full phase covariance of both the singular and minimal models and develops the corresponding inference problem at the observational level.
The present work is concerned with the same question at the observational level. Radio timing does not measure angular velocities. It measures pulse times of arrival, which are converted into rotational phases and then into timing residuals by fitting and subtracting a deterministic timing model that accounts for the spin frequency, the spin-down and other deterministic effects \citep{Hobbs:2006cd, Edwards:2006zg, Manchester:2015mda}. Only the crust participates in this, since the pulses are locked to the crust. The data that are available to test the physics are therefore crust-only phase residuals from which the lowest-order trends have been removed, and a test of spin noise stationarity must be carried out on these data, or on spin-frequency time series derived from them by differentiation \citep{ONeill:2024uiw}.

Each step between the physical state and the analyzed data has consequences that must be understood before the physics can be read off. Integration of a stationary angular velocity produces a phase that is not stationary, so that the nonstationarity of the phase is not by itself evidence for a nonstationary spin. The fitting and removal of the timing model suppresses the lowest frequencies, where the difference between a random walk and a stationary process is largest, and is itself known to leave nonstationary residuals \citep{Lee:2012bf, vanHaasteren:2012hj, Allen:2025waa}. Differentiating the phases back into spin frequencies averages the state over finite windows and thins the sampling. None of these steps is the object of this work, but each of them could bias a test of stationarity that is carried out at the observational level, and each must therefore be treated exactly. We do so by deriving the phase-domain likelihood of the two models building on the analytical solutions of \citet{Bernardo:2026opv}, with the timing model marginalized analytically, and by comparing inference in phase with inference in derived spin frequencies on the same simulated data. The result is a scalable, exact Gaussian process method for the phases,
and an assessment of what radio timing alone can and cannot establish about the stationarity of spin noise.

We first present the exact phase means and covariances of the singular and minimal models (Section~\ref{sec:phase_moments}). We then derive the phase likelihood with linear scaling in the number of observations, the analytic marginalization of the timing model, and the evidence and superfluid-prediction machinery used throughout (Section~\ref{sec:likelihood}). The method is tested on simulated phase data generated from the singular and the minimal model, with full-state and crust-only observations (Sections~\ref{sec:singular_simulations}--\ref{sec:minimal_simulations}). We then ask whether the same conclusions are reached when the phases are first processed into local spin frequencies (Section~\ref{sec:phase_or_angular_velocity}), and we discuss what pulsar timing can and cannot do to test the stationarity of spin noise and the internal conservation of angular momentum (Section~\ref{sec:testing_conservation}). We conclude in Section~\ref{sec:conclusions}.

We include four appendices with additional details.  Appendix~\ref{app:kernels} derives the finite-start kernel orientation and its multi-output symmetry; Appendix~\ref{app:ss} presents the semiseparable factorization and the forward and backward solves that give the likelihood its linear cost; Appendix~\ref{app:verification} documents the analytic and numerical checks used to validate the implementation; and Appendix~\ref{app:evidence} records the evidence settings and reproducibility metadata for the cached ensembles.

\section{Exact phase means and covariances}
\label{sec:phase_moments}

We present the phase means and covariances for both the singular (Section \ref{subsec:singular}) and the minimal (Section \ref{subsec:minimal}) models. 

\subsection{Singular nonstationary-spin model}
\label{subsec:singular}

For the singular model \citep{Meyers:2021myb, Meyers:2021slh}, we write down the component equations
\begin{align}
 I_{\rm c}\dot\Omega_{\rm c} &=-\frac{I_{\rm c}}{\tau_{\rm c}}(\Omega_{\rm c}-\Omega_{\rm s})+N_{\rm c}+\xi_{\rm c} \,,\\
 I_{\rm s}\dot\Omega_{\rm s} &= \frac{I_{\rm s}}{\tau_{\rm s}}(\Omega_{\rm c}-\Omega_{\rm s})+N_{\rm s}+\xi_{\rm s} \,,
\end{align}
where the dot over a variable denotes a time derivative and the $\xi_i(t)$'s are stochastic torques with a zero mean and a covariance $\langle\xi_i(t)\xi_j(t')\rangle=I_i^2Q_i\delta_{ij}\delta(t-t')$.
We work throughout in the detailed-balance limit $I_{\rm s}/I_{\rm c}=\tau_{\rm s}/\tau_{\rm c}$, in which the mutual-friction torques balance each other.
The relaxation time and ratio are
\begin{align}
\tau &=\frac{\tau_{\rm c}\tau_{\rm s}}{\tau_{\rm c}+\tau_{\rm s}} \,, \\
r &=\frac{\tau_{\rm s}}{\tau_{\rm c}} \,,
\end{align}
so that $(\tau, r, Q_{\rm c}, Q_{\rm s})$ are the four parameters of spin wandering, with $r=x_{\rm s}/x_{\rm c}$ in terms of the fractional moments of inertia $x_i=I_i/(I_{\rm c}+I_{\rm s})$ in this limit. For elapsed times $u=t-t_0$, $u'=t'-t_0$ measured from the start $t_0$ of the observations, we define $u_{<}=\min(u,u')$, $u_{>}=\max(u,u')$ and $\Delta=|u-u'|$.

The means are compact in the eigenmode basis.  With
\begin{align}
 \mathcal M_+(u)&=\Omega_{+,0}u+\frac{N_+}{2I}u^2 \,,\\
 \mathcal M_-(u)&=\Omega_{-,\infty}u+\tau(\Omega_{-,0}-\Omega_{-,\infty})(1-\e^{-u/\tau}) \,,
\end{align}
we obtain
\begin{align}
 \langle\phi_{\rm c}\rangle&=\phi_{{\rm c},0}+\frac{\tau}{T_\Sigma}\mathcal M_+ +\frac{\tau_{\rm s}}{T_\Sigma}\mathcal M_- \,,\\
 \langle\phi_{\rm s}\rangle&=\phi_{{\rm s},0}+\frac{\tau}{T_\Sigma}\mathcal M_+ -\frac{\tau_{\rm c}}{T_\Sigma}\mathcal M_- \,,
\end{align}
where $T_\Sigma=\tau_{\rm c}+\tau_{\rm s}$, $I=I_{\rm c}+I_{\rm s}$, $\Omega_- = \Omega_{\rm c} - \Omega_{\rm s}$ is the crust--superfluid lag and $\Omega_+ = (\tau_{\rm c} \Omega_{\rm c} + \tau_{\rm s} \Omega_{\rm s})/\tau$ is the diffusive mode, which is proportional to the total angular momentum, $J=(\tau/T_\Sigma)I\Omega_+$, in the detailed-balance limit. The torque $N_+=(I/\tau)(\tau_{\rm c}N_{\rm c}/I_{\rm c}+\tau_{\rm s}N_{\rm s}/I_{\rm s})$ is the external torque projected on the diffusive mode, such that $\langle\dot\Omega_+\rangle=N_+/I$, and $\Omega_{-,\infty}=\tau(N_{\rm c}/I_{\rm c}-N_{\rm s}/I_{\rm s})$ is the steady-state lag. The remaining quantities $(\Omega_{+,0}, \Omega_{-,0}, \phi_{\rm c, 0}, \phi_{\rm s, 0})$ are the values of the modes and phases at $t_0$, which are to be drawn from an initial distribution. The constant, linear and quadratic pieces in $\langle\phi_{\rm c}\rangle$ correspond to a phase offset, a spin-frequency offset and spin-down, respectively. In the likelihood we marginalize over these long-term trends, taking them into the timing model (Section~\ref{sec:likelihood}). Only the exponentially relaxing piece of $\mathcal M_-$ is not absorbed by the timing model; it decays within a few $\tau$ of the start of the observations and vanishes identically when the lag is initialized at its steady state, $\Omega_{-,0}=\Omega_{-,\infty}$.

The four scalar kernels required by the exact phase covariance are
\begin{align}
 \mathcal I_1(u,u')&=\frac{u_<^2(3u_>-u_<)}{6},\label{eq:I1}\\
 \mathcal I_3(u,u')&=u_<-\tau+\tau\e^{-u/\tau}+\tau\e^{-u'/\tau}
 -\frac{\tau}{2}\e^{-\Delta/\tau}-\frac{\tau}{2}\e^{-(u+u')/\tau},\label{eq:I3}\\
 \mathcal I_{+-}(u,u')&=uu_<-\frac{u_<^2}{2}
 -\tau(u-u_<+\tau)\e^{-(u'-u_<)/\tau}
 \nonumber \\
 & \qquad +\tau(u+\tau)\e^{-u'/\tau},\label{eq:Ipm}\\
 \mathcal I_{-+}(u,u')&=\mathcal I_{+-}(u',u) \,,
\end{align}
We define
\begin{align}
 \mathcal I_2 (u,u')&\equiv\mathcal I_{+-}(u,u') +\mathcal I_{-+}(u,u') \,, \nonumber \\
&= (u_<-\tau)(u_>+\tau)-\tau^2\e^{-\Delta/\tau}
 +\tau(u+\tau)\e^{-u'/\tau} \nonumber \\
 & \qquad +\tau(u'+\tau)\e^{-u/\tau} \,. \label{eq:I2}
\end{align}
The kernels ${\cal I}_1(t,t')$, ${\cal I}_2(t,t')$, and ${\cal I}_3(t,t')$ were derived in \cite{Bernardo:2026opv} to express the phase covariance of the singular model; they are the double time integrals of the diffusive, cross and damped pieces of the angular velocity covariance, respectively, and Appendix~\ref{app:kernels} summarizes their derivation.

We define transformed noise intensities
\begin{align}
 \tau^2Q_+ &=\tau_{\rm c}^2Q_{\rm c}+\tau_{\rm s}^2Q_{\rm s} \,, \\ 
 Q_- &=Q_{\rm c}+Q_{\rm s} \,, \\
 \tau Q_\times &=\tau_{\rm c}Q_{\rm c}-\tau_{\rm s}Q_{\rm s} \,.
\end{align}
The phase covariance blocks are
\begin{align}
 K_{\rm cc}&=\frac{\tau^2}{T_\Sigma^2}
 [Q_+\mathcal I_1+Q_\times\tau_{\rm s}\mathcal I_2+Q_-\tau_{\rm s}^2\mathcal I_3] \,,\label{eq:Kcc}\\
 K_{\rm ss}&=\frac{\tau^2}{T_\Sigma^2}
 [Q_+\mathcal I_1-Q_\times\tau_{\rm c}\mathcal I_2+Q_-\tau_{\rm c}^2\mathcal I_3] \,,\label{eq:Kss}\\
 K_{\rm cs}&=\frac{\tau^2}{T_\Sigma^2}\{Q_+\mathcal I_1-Q_-\tau_{\rm c}\tau_{\rm s}\mathcal I_3
 +Q_\times[\tau_{\rm s}\mathcal I_{-+}-\tau_{\rm c}\mathcal I_{+-}]\} \,,\label{eq:Kcs}\\
 K_{\rm sc}&=\frac{\tau^2}{T_\Sigma^2}\{Q_+\mathcal I_1-Q_-\tau_{\rm c}\tau_{\rm s}\mathcal I_3
 +Q_\times[\tau_{\rm s}\mathcal I_{+-}-\tau_{\rm c}\mathcal I_{-+}]\} \,.\label{eq:Ksc}
\end{align}
Then, $K_{\rm sc}(t,t')=K_{\rm cs}(t',t)$, but $K_{\rm cs}$ need not be symmetric under the exchange of times alone.
At late times compared to the initial time, the leading equal-time variance is $K_{ii}\sim \tau^2Q_+u^3/(3T_\Sigma^2)$, exposing the diffusive total angular momentum mode.\footnote{In general, the diffusive mode $\Omega_+$ is not the total angular momentum mode $J$. We restrict our attention to the detailed balance limit where the mutual friction torques balance and, as a consequence, the connection
$J = (\tau/T_\Sigma) I\Omega_+$
appears.}
The $u^3$ growth is the phase-level signature of a random walk in the angular velocity. It is the double time integral of the linearly growing variance of $\Omega_+$.

\subsection{Minimal stationary-spin model}
\label{subsec:minimal}

The minimal model is obtained by requiring the stochastic torques to be equal and opposite, $\xi_{\rm c}(t)=-\xi_{\rm s}(t)\equiv-\xi(t)$, so that the total angular momentum receives no stochastic forcing \citep{Meyers:2021myb,Antonelli:2022gqw}. The component equations are
\begin{align}
 I_{\rm c}\dot\Omega_{\rm c} &=-\frac{I_{\rm c}I_{\rm s}}{I}\frac{(\Omega_{\rm c}-\Omega_{\rm s})}{\tau}+N_{\rm c}+\xi \,,\\
 I_{\rm s}\dot\Omega_{\rm s} &= \frac{I_{\rm c}I_{\rm s}}{I}\frac{(\Omega_{\rm c}-\Omega_{\rm s})}{\tau}-\xi \,,
\end{align}
with 
$\langle \xi(t) \rangle = 0$,
$\langle\xi(t)\xi(t')\rangle=I^2Q\,\delta(t-t')$ and $I=I_{\rm c}+I_{\rm s}$ as before. Writing $x_i=I_i/I$ for the fractional moments of inertia, the eigenmode form of the equations is
\begin{align}
 \dot J &= N_{\rm c} \,,\\
 \dot\Omega_- &= -\frac{\Omega_-}{\tau}+\frac{N_{\rm c}}{I_{\rm c}}+\frac{\xi}{x_{\rm c}x_{\rm s}I} \,.
\end{align}
The total angular momentum evolves deterministically and the lag is an Ornstein--Uhlenbeck (OU) process, the only random dynamical variable. The stochastic residuals of the components are tied to the lag through
\begin{equation}
 \delta\Omega_{\rm c}=x_{\rm s}\,\delta\Omega_- \,,\qquad
 \delta\Omega_{\rm s}=-x_{\rm c}\,\delta\Omega_- \,,
\end{equation}
which is the differential form of the conservation law $x_{\rm c}\delta\Omega_{\rm c}+x_{\rm s}\delta\Omega_{\rm s}=0$. The model has three spin-wandering parameters, $(\tau, x_{\rm s}, Q)$, compared to four for the singular model.

The deterministic parts of the minimal and singular equations of motion coincide in the detailed-balance limit, since $I_{\rm c}/\tau_{\rm c}=I_{\rm c}x_{\rm s}/\tau$ and $I_{\rm s}/\tau_{\rm s}=I_{\rm s}x_{\rm c}/\tau$. The phase means are therefore given by the same expressions as in Section~\ref{subsec:singular}, with $N_{\rm s}=0$, $\tau_{\rm s}/T_\Sigma=x_{\rm s}$ and $\tau_{\rm c}/T_\Sigma=x_{\rm c}$:
\begin{align}
 \langle\phi_{\rm c}\rangle&=\phi_{{\rm c},0}+x_{\rm c}x_{\rm s}\mathcal M_+ + x_{\rm s}\mathcal M_- \,,\\
 \langle\phi_{\rm s}\rangle&=\phi_{{\rm s},0}+x_{\rm c}x_{\rm s}\mathcal M_+ - x_{\rm c}\mathcal M_- \,,
\end{align}
with $N_+=N_{\rm c}/(x_{\rm c}x_{\rm s})$ and $\Omega_{-,\infty}=\tau N_{\rm c}/I_{\rm c}$. As in the singular case, the constant, linear and quadratic pieces are absorbed by the timing model. The two models thus share their deterministic evolution and differ only in the statistics of their fluctuations, which is the reason that the phase covariance, rather than the mean, carries the distinction between them.

Because the lag is stationary, it is natural to initialize it from its stationary distribution, $\delta\Omega_-(t_0)\sim\mathcal N(0, Q\tau/(2x_{\rm c}^2x_{\rm s}^2))$, rather than from a fixed value; this is the initialization adopted throughout this work. For stationary lag initialization, the phase covariance is built from the single kernel
\begin{equation}
 \mathcal H(u,u')=2\tau u_<-\tau^2[1+\e^{-\Delta/\tau}-\e^{-u/\tau}-\e^{-u'/\tau}] \,, \label{eq:H}
\end{equation}
which is the double time integral of the stationary OU covariance $\e^{-\Delta/\tau}$, and reads
\begin{equation}
 \bK_{\min}(t,t')=\frac{Q\tau}{2}\mathcal H(u,u')
 \begin{pmatrix}
 x_{\rm c}^{-2}&-(x_{\rm c}x_{\rm s})^{-1}\\
 -(x_{\rm c}x_{\rm s})^{-1}&x_{\rm s}^{-2}
 \end{pmatrix} \,.\label{eq:Kminimal}
\end{equation}
Note that this cannot be obtained with a $Q_+=0$ substitution into the singular model, which would correspond to a lag fixed at $t_0$. The stochastic initialization of the stationary lag contributes an additional separable term. Indeed,
\begin{equation}
 \mathcal H=2\tau\mathcal I_3+\tau^2(1-\e^{-u/\tau})(1-\e^{-u'/\tau}) \,,
\end{equation}
where the second term is the variance of the initial lag propagated to the phases. The crust block is conveniently written as
\begin{equation}
 K_{\rm cc}(t,t')=\frac{A_{\rm min}}{2}\mathcal H(u,u') \,,\qquad A_{\rm min}\equiv\frac{Q\tau}{x_{\rm c}^2} \,,\label{eq:Amin}
\end{equation}
where $A_{\rm min}$ is the stationary variance of the crust angular velocity up to a factor of two. The crust phase covariance therefore depends on the two combinations $(\tau, A_{\rm min})$ only; the inertia fraction $x_{\rm s}$ enters exclusively through the superfluid and cross blocks, $K_{\rm ss}=(x_{\rm c}/x_{\rm s})^2K_{\rm cc}$ and $K_{\rm cs}=K_{\rm sc}=-(x_{\rm c}/x_{\rm s})K_{\rm cc}$. This observation controls much of what follows. Crust-only data cannot inform $x_{\rm s}$ in the minimal model.

The angular velocities are stationary while $K_{\rm cc}(t,t)\propto u$ for $u\gg\tau$.  Integration-induced phase nonstationarity persists in a model whose stochastic spin/angular velocity is stationary \citep{Antonelli:2022gqw, Bernardo:2026opv}. The conservation law
\begin{equation}
\label{eq:minimal_eye}
 x_{\rm c}\delta\phi_{\rm c}+x_{\rm s}\delta\phi_{\rm s}=0
\end{equation}
makes each component-time covariance block rank one and turns an external measurement of $x_{\rm s}$ into a definite hidden-state prediction.

\section{Linear-scaling phase likelihood}
\label{sec:likelihood}

We model the phase observed at epochs $t_1<\dots<t_N$ as
\begin{equation}
 y_{i,n}=\delta\phi_i(t_n)+\sum_{\mu=0}^{2}X_{n\mu}\beta_{i,\mu}+\varepsilon_{i,n} \,,\qquad i\in\{{\rm c},{\rm s}\} \,,\label{eq:measurement}
\end{equation}
where $\delta\phi_i$ is the zero-mean stochastic phase whose covariance was derived in Section~\ref{sec:phase_moments}, $\varepsilon_{i,n}$ is Gaussian measurement noise with variance $\sigma_{i,n}^2$, independent between epochs and components, and $X_{n\mu}\beta_{i,\mu}$ is the deterministic timing model. The columns of the $N\times3$ design matrix
\begin{equation}
 \bX=\begin{pmatrix}1&t_1-t_{\rm ref}&(t_1-t_{\rm ref})^2/2\\ \vdots&\vdots&\vdots\\1&t_N-t_{\rm ref}&(t_N-t_{\rm ref})^2/2\end{pmatrix}
\end{equation}
carry the phase offset, the spin-frequency offset and the spin-down of each component, so that $\bbeta_i=(\beta_{i,0},\beta_{i,1},\beta_{i,2})$ collects the constant, linear and quadratic pieces of the phase means of Section~\ref{sec:phase_moments}, which we do not attempt to predict but instead marginalize. In radio timing the crust phase is the observable, and timing residuals $r(t)$ relative to a timing model are related to it through $\delta\phi_{\rm c}\simeq2\pi\nu\, r$ for a nominal spin frequency $\nu$ \citep{ONeill:2024uiw, Dong:2025nho, 2025arXiv251011077K}.

For full-state observations, which we use in the simulations as an identifiability benchmark, the data vector $\by=(y_{{\rm c},1},\dots,y_{{\rm c},N},y_{{\rm s},1},\dots,y_{{\rm s},N})$ has length $2N$, the covariance is the $2N\times2N$ block matrix
\begin{equation}
 \bK=\begin{pmatrix}K_{\rm cc}&K_{\rm cs}\\K_{\rm sc}&K_{\rm ss}\end{pmatrix}+\diag(\sigma_{{\rm c},n}^2,\sigma_{{\rm s},n}^2) \,,
\end{equation}
with the oriented blocks of (\ref{eq:Kcc}-\ref{eq:Ksc}) or \eqref{eq:Kminimal}, and the timing design is $\bX\oplus\bX$, one quadratic per component.
Each block of this multi-output covariance is a sum of polynomial-exponential terms in the two times, so the full-state matrix also admits a block semiseparable representation,
and a linear-scaling factorization of the same type as the one described below.
Full-state phase data are a hypothetical observable that serves here as a benchmark for what the crust alone can and cannot establish, the simulated data sets are small, and the full-state likelihood is therefore evaluated by a dense Cholesky factorization of the $2N\times2N$ matrix at ${\cal O}(N^3)$ cost. Our effort is directed at the crust-only likelihood, which is the one that applies to radio timing data and whose length is set by the observations.

\subsection*{Semiseparable crust-only likelihood}

The observationally relevant likelihood is that of the crust alone, for which $\by$ has length $N$ and $\bK=K_{\rm cc}+\diag(\sigma_n^2)$. Exact evaluation at realistic cadences requires a representation of $K_{\rm cc}$ that avoids forming or inverting a dense $N\times N$ matrix. At ordered epochs $t_i$, both models' crust covariance matrices have the semiseparable form
\begin{equation}
 K_{ij}=a_i\delta_{ij}+\begin{cases}
 \boldsymbol U_i\boldsymbol V_j^{\mathsf T} \,, &i>j \,,\\
 \boldsymbol V_i\boldsymbol U_j^{\mathsf T} \,, &i<j \,.
 \end{cases}\label{eq:ssform}
\end{equation}
The singular phase kernel has rank three, with a convenient basis proportional to $(\e^{-u/\tau},u,1)$; the independently derived minimal kernel has rank two, proportional to $(\e^{-u/\tau},1)$.  Measurement variances enter only through $a_i$, which is what makes the semiseparable representation compatible with heteroscedastic noise. The explicit factors $\boldsymbol U_i$, $\boldsymbol V_j$ and $a_i$ are collected in Appendix~\ref{app:ss}.

For rank $J$, the Cholesky recurrences require $\mathcal O(NJ^2)$ operations and $\mathcal O(NJ)$ storage.  Starting with $\boldsymbol S_0=0$, one step is
\begin{align}
 d_i&=a_i-\boldsymbol U_i\boldsymbol S_i\boldsymbol U_i^{\mathsf T} \,,\\
 \boldsymbol W_i&=\frac{\boldsymbol V_i-\boldsymbol U_i\boldsymbol S_i}{d_i} \,,\\
 \boldsymbol S_{i+1}&=\boldsymbol S_i+d_i\boldsymbol W_i^{\mathsf T}\boldsymbol W_i \,.
\end{align}
The recurrences produce the factorization $\bK=\boldsymbol L\boldsymbol D\boldsymbol L^{\mathsf T}$ with $\boldsymbol D=\diag(d_1,\dots,d_N)$ and a unit lower-triangular $\boldsymbol L$ whose strictly lower part is itself semiseparable, $L_{ij}=\boldsymbol U_i\boldsymbol W_j^{\mathsf T}$ for $i>j$. The log-determinant is a sum of $N$ terms,
\begin{equation}
 \log|\bK|=\sum_{i=1}^N\log d_i \,,\label{eq:logdet_ss}
\end{equation}
and the action of the inverse on a vector $\boldsymbol b$ is obtained without ever forming $\bK^{-1}$,
\begin{equation}
 \bK^{-1}\boldsymbol b=\boldsymbol L^{-\mathsf T}\boldsymbol D^{-1}\boldsymbol L^{-1}\boldsymbol b \,,\label{eq:kinv_ss}
\end{equation}
by a forward pass that computes $\boldsymbol z=\boldsymbol D^{-1/2}\boldsymbol L^{-1}\boldsymbol b$ and a backward pass that computes $\boldsymbol L^{-\mathsf T}\boldsymbol D^{-1/2}\boldsymbol z$. Each pass carries a $J$-dimensional accumulator and visits every epoch once (Appendix~\ref{app:ss}), so a solve costs ${\cal O}(NJ)$. The quadratic form needed by the likelihood is then simply $\boldsymbol b^{\mathsf T}\bK^{-1}\boldsymbol b=\sum_i z_i^2$. Since $J$ is fixed by the physics of the two-component model, the factorization, the log-determinant and every solve scale linearly with the number of observations, in contrast to the ${\cal O}(N^3)$ cost and ${\cal O}(N^2)$ storage of a dense factorization.
The construction is inspired by \cite{2015ITPAM..38..252A, 2017AJ....154..220F} for stationary scalar kernels; here the factors are fixed by the finite-start, nonstationary phase covariance of a hidden two-component system rather than by a phenomenological kernel.

\subsection*{Quadratic timing-trend marginalization}

The timing-model coefficients $\bbeta$ of \eqref{eq:measurement} are nuisance parameters.
They absorb the deterministic spin-down together with the phase and spin-frequency offsets of the phase means,
and in this sense they are not separate from the physics. In the singular model the quadratic trend is the deterministic part of the same diffusive mode whose fluctuations render the spin nonstationary, whereas in the minimal model the quadratic trend is the deterministic evolution of a conserved total angular momentum and the fluctuations about it are stationary. What the timing model removes is the same in both cases, so that the comparison between the models rests on the fluctuations that remain. With a flat prior on $\bbeta$, they can be integrated out analytically, and the exact marginalized likelihood for the spin-noise parameters $\theta$ is
\begin{align}
 & \log p(\by|\theta) \nonumber \\
 & =-\frac12\left[
 \boldsymbol r^{\mathsf T}\bK^{-1}\boldsymbol r+\log|\bK|
 +\log|\bX^{\mathsf T}\bK^{-1}\bX|+(N-3)\log(2\pi)
 \right] \,, \label{eq:gls}
\end{align}
where
\begin{equation}
 \widehat{\bbeta}=(\bX^{\mathsf T}\bK^{-1}\bX)^{-1}\bX^{\mathsf T}\bK^{-1}\by,
 \qquad \boldsymbol r=\by-\bX\widehat{\bbeta}
\end{equation}
are the generalized least-squares (GLS) estimate of the trend at fixed $\theta$ and the corresponding residual. All appearances of $\bK^{-1}$ in \eqref{eq:gls} are inverse actions on $\by$ and on the three columns of $\bX$, computed with the ${\cal O}(N)$ solve of \eqref{eq:kinv_ss}, so that one likelihood evaluation costs a single factorization and four solves. The determinant of the normal matrix is part of the normalization; omitting it changes the evidence. In the full-state case the same expression holds with $\bX\to\bX\oplus\bX$ and $N-3\to2N-6$.

The ordinary least-squares (OLS) residual maker
\begin{equation}
 \boldsymbol P=\boldsymbol I-\bX(\bX^{\mathsf T}\bX)^{-1}\bX^{\mathsf T}
\end{equation}
is used only to display all models and truths in one parameter-independent space.  It is not substituted for the parameter-dependent GLS marginalization in \eqref{eq:gls}.

For a sinusoidal probe $\boldsymbol e_f$ on an irregular cadence, the empirical transmission is $\|\boldsymbol P\boldsymbol e_f\|^2/\|\boldsymbol e_f\|^2$.  Removing a quadratic phase model yields an $f^6$ response at $f\ll1/T$; removing a linear frequency trend yields $f^4$ \citep{Allen:2025waa, Pitrou:2024scp, Hazboun:2019vhv}.
The projection reduces the total power of the phase residuals, and it does so mostly by suppressing the power at the lowest frequencies, where the spectrum of a red process is largest. This is why the projected phases in Figure~\ref{fig:paths} have much smaller excursions than the raw ones and why their appearance is dominated by fluctuations on time scales shorter than the observation span.

\subsection*{Simulated phase data}
\label{subsec:simulated_data}

\begin{figure*}
 \centering
 \includegraphics[width=0.49\textwidth]{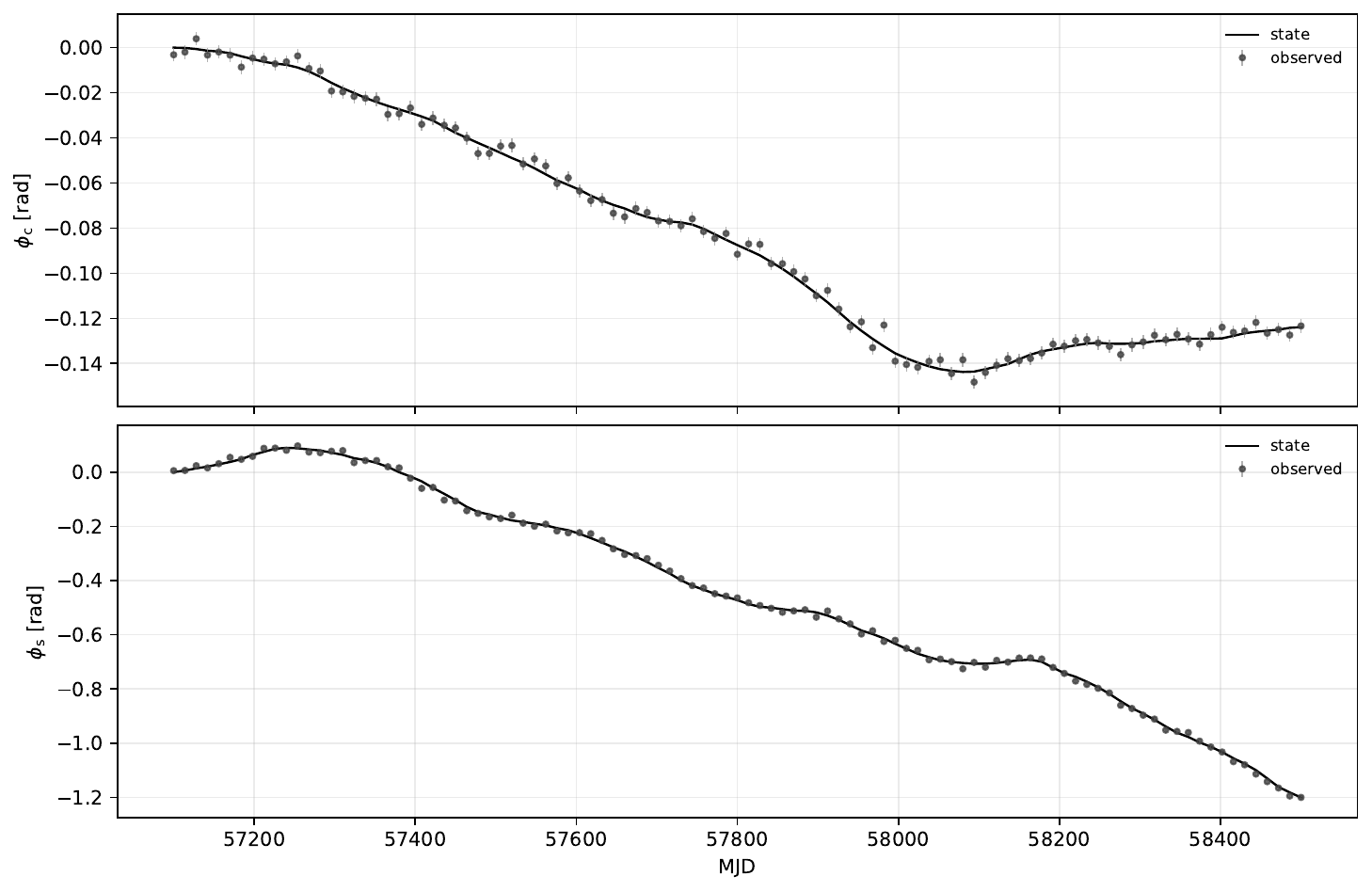}\hfill
 \includegraphics[width=0.49\textwidth]{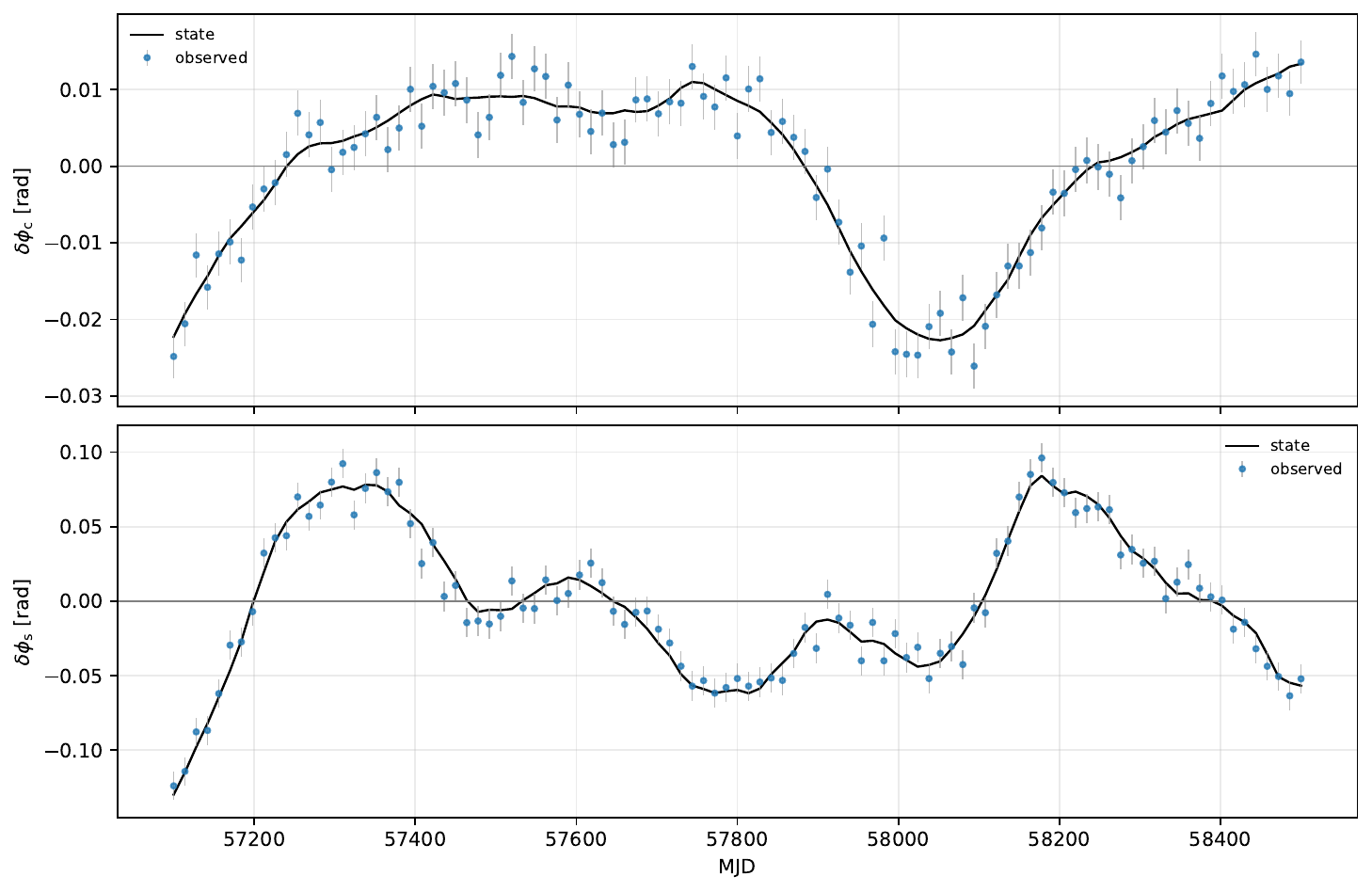}
 \includegraphics[width=0.49\textwidth]{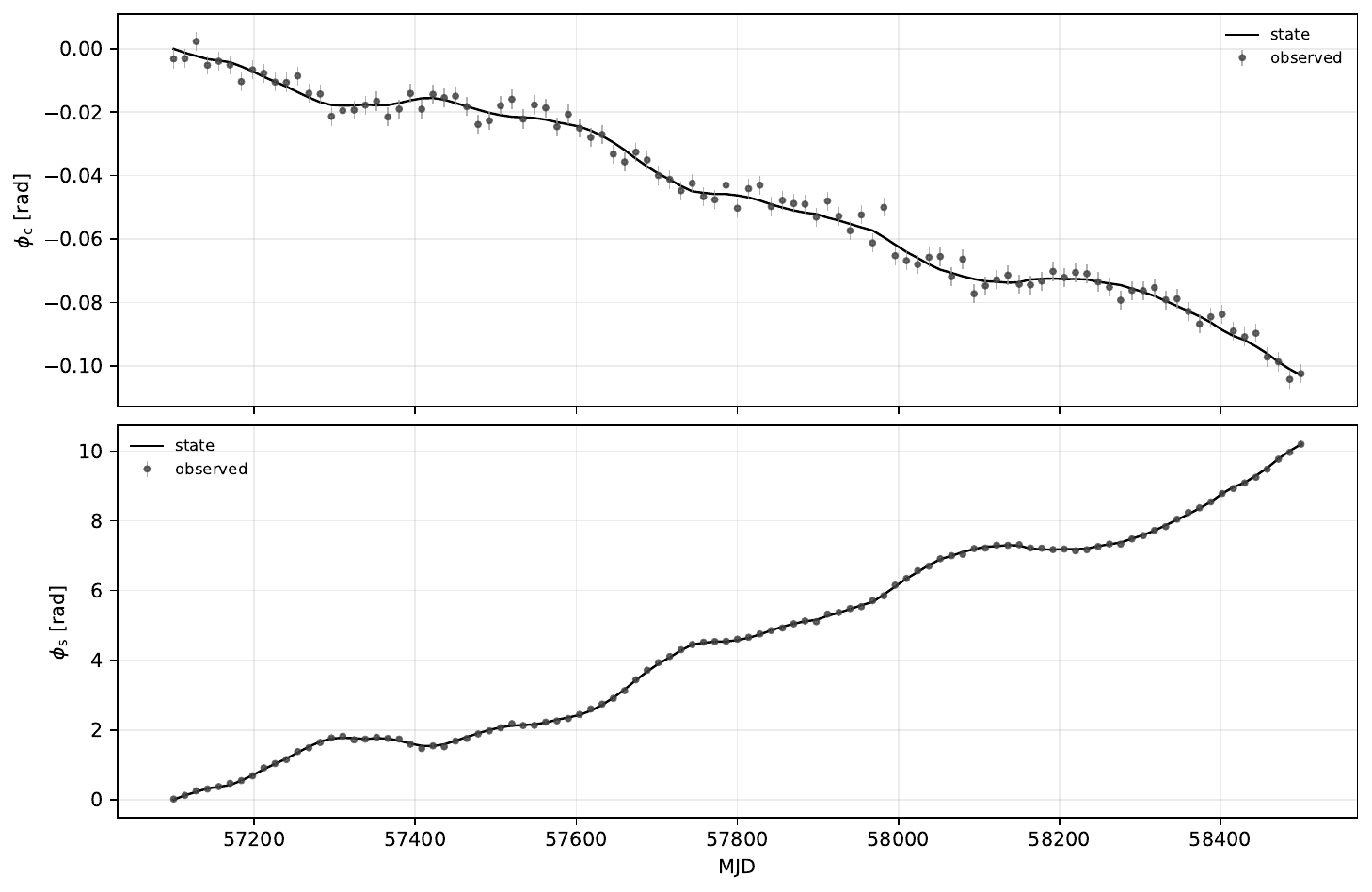}\hfill
 \includegraphics[width=0.49\textwidth]{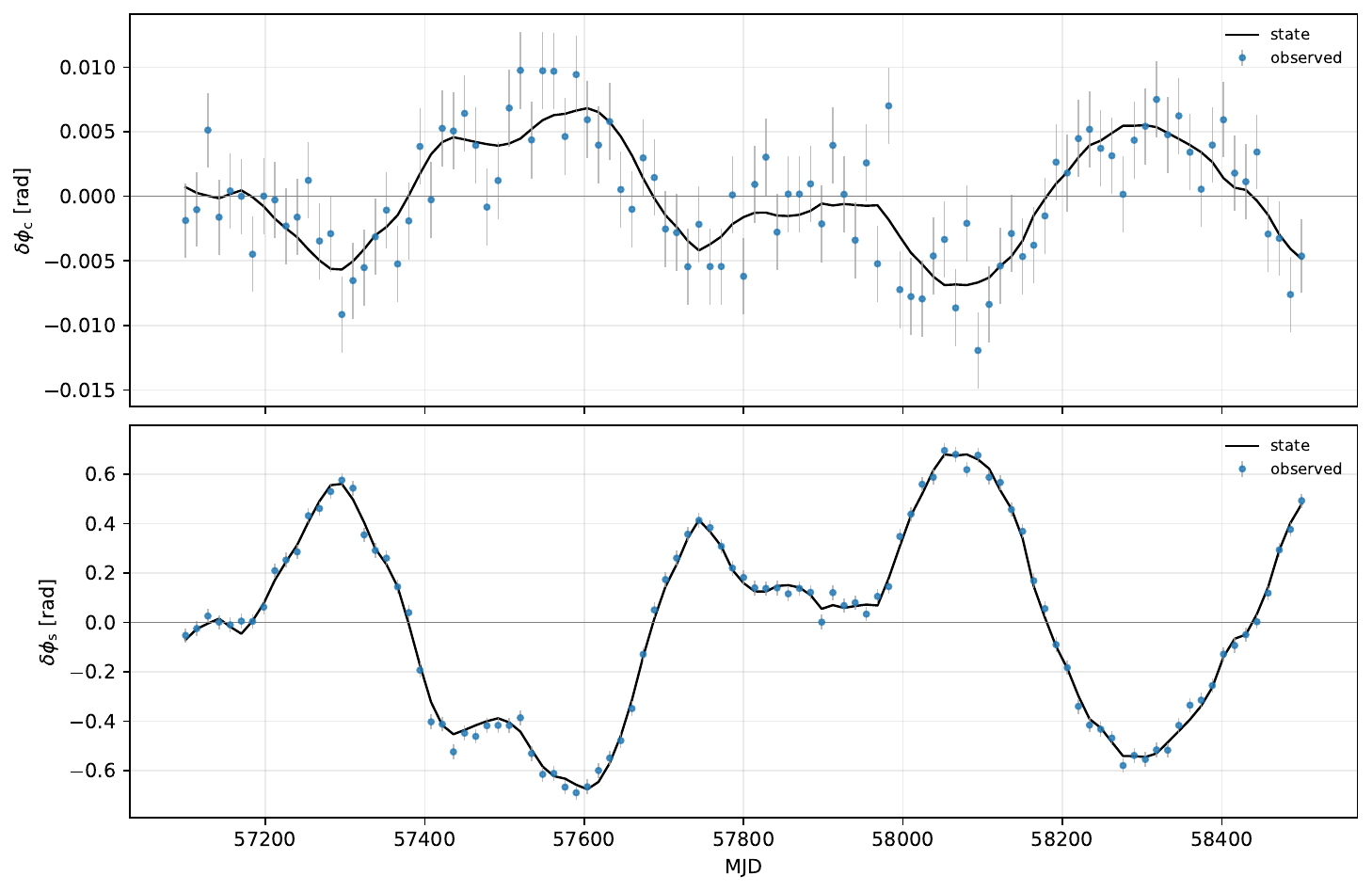}
 \caption{
Phase residual simulation for both singular (top) and minimal (bottom) models before (left) and after (right) quadratic projection. The black curves are the latent crust and superfluid phases drawn exactly from the analytical covariances, and the points with error bars are the observations, obtained by adding independent Gaussian measurement noise with standard deviations $\sigma_{\rm c}=3\times10^{-3}$ rad for the crust and $\sigma_{\rm s}=10^{-2}$ rad (singular) or $3\times10^{-2}$ rad (minimal) for the superfluid. The model parameters are given in Table \ref{tab:simulation_parameters}. The same OLS projection is applied to the latent phases and to the observations.
}
\label{fig:paths}
\end{figure*}

Figure \ref{fig:paths} shows simulated noisy singular and minimal phase paths before and after projection. The simulation makes direct use of the analytical solutions of Section~\ref{sec:phase_moments}. Because the models are linear and the stochastic torques are Gaussian, the crust and superfluid phases at the $N$ observing epochs are jointly a $2N$-dimensional Gaussian random vector whose mean and covariance are given exactly by the phase means and by the block covariance of (\ref{eq:Kcc}--\ref{eq:Ksc}) or \eqref{eq:Kminimal}. A realization is a single draw from this multivariate normal distribution,
\begin{equation}
 \delta\boldsymbol\phi\sim\mathcal N\!\left(\langle\boldsymbol\phi\rangle,\,\bK\right) \,,
\end{equation}
which we obtain in practice by multiplying a vector of independent standard normal variates by a matrix square root of $\bK$. Since the analytical moments are exact for arbitrary sampling times, no time stepping of the stochastic differential equations is involved, and the realizations reproduce the finite-start and stationary-lag structure of the kernels without any discretization error. The deterministic means are set to zero, since their constant, linear and quadratic pieces are marginalized in the analysis, and the minimal lag is initialized from its stationary distribution (Section~\ref{subsec:minimal}). We generate $N=101$ evenly sampled epochs over 1400 days with a cadence of 14 days.

Measurement noise is added as independent Gaussian variates, uncorrelated between epochs and between the crust and superfluid channels, with standard deviations $\sigma_{\rm c}$ and $\sigma_{\rm s}$ listed in Table~\ref{tab:simulation_parameters}. The crust noise level is chosen to be a sizeable fraction of the projected crust excursions, so that the error bars are visible in the projected data while the stochastic phase remains clearly detected; the superfluid noise is taken larger in absolute terms, since the superfluid channel is a hypothetical observable, but it remains small relative to the superfluid excursions, which exceed those of the crust for the parameters of Table~\ref{tab:simulation_parameters} (by the factor $x_{\rm c}/x_{\rm s}$ in the minimal model).
Their orders of magnitude are broadly motivated by the spin-noise parameters inferred for PSR J1359$-$6038 \citep{ONeill:2024uiw}, rounded and rescaled so that the observation window contains several relaxation times ($\tau$ of one year) and the realizations are visibly noisy. The minimal amplitude $A_{\rm min}$ was chosen to give projected crust excursions comparable to the singular case over the observation window.
It is worth noting that, when both components are shown, the two realizations in Figure~\ref{fig:paths} can be told apart by eye even after projection. In the minimal realization the crust and superfluid phases are mirror images of each other up to the factor $-x_{\rm c}/x_{\rm s}$, as required by the conservation law \eqref{eq:minimal_eye}, whereas in the singular realization the two components share a common wander and no such proportionality holds. =
With the crust alone, the projected phases of the two realizations look alike, and telling the models apart becomes the quantitative problem that the rest of this work addresses.

\begin{table}
    \centering
    \small
    \begingroup
    \renewcommand{\arraystretch}{1.00}
    \setlength{\tabcolsep}{4pt}
    \caption{Reference parameters for the singular and minimal models and the phase measurement noise considered throughout this work. Both simulations use $N=101$ evenly sampled epochs with a cadence of 14 days over 1400 days.}
    \begin{tabular}{@{}lcc@{}}
        \hline
        Parameter & Value & Unit \\
        \hline\hline
        \multicolumn{3}{@{}l}{\textit{Singular-nonstationary}} \\
        $\tau$ & $365$ & day \\
        $r$ & $0.1$ & -- \\
        $Q_{\rm c}$ & $10^{-10}$ & rad$^2$ day$^{-3}$ \\
        $Q_{\rm s}$ & $10^{-8}$ & rad$^2$ day$^{-3}$ \\
        $\sigma_{\rm c}$ & $3\times10^{-3}$ & rad \\
        $\sigma_{\rm s}$ & $10^{-2}$ & rad \\
        \hline\hline
        \multicolumn{3}{@{}l}{\textit{Minimal-stationary}} \\
        $\tau$ & $365$ & day \\
        $A_{\rm min}$ & $3\times10^{-8}$ &
            rad$^2$ day$^{-2}$ \\
        $x_{\rm s}$ & $10^{-2}$ & -- \\
        $\sigma_{\rm c}$ & $3\times10^{-3}$ & rad \\
        $\sigma_{\rm s}$ & $3\times10^{-2}$ & rad \\
        \hline
    \end{tabular}
    \label{tab:simulation_parameters}
    \endgroup
\end{table}

\subsection*{Bayesian evidence and model comparison}
\label{subsec:evidence}

The question of whether the spin noise is stationary is posed as a comparison between the minimal and singular models given the same phase data. For a model $\mathcal M$ with spin-noise parameters $\theta$ and prior $\pi(\theta|\mathcal M)$, the evidence is the trend-marginalized likelihood of \eqref{eq:gls} integrated over the prior,
\begin{equation}
 Z_{\mathcal M}=p(\by|\mathcal M)=\int\dd\theta\, p(\by|\theta,\mathcal M)\,\pi(\theta|\mathcal M) \,,
\end{equation}
and the models are compared through the log Bayes factor
\begin{equation}
 \ln B_{\rm singular,minimal}=\ln Z_{\rm singular}-\ln Z_{\rm minimal} \,,\label{eq:bayes_factor}
\end{equation}
so that positive values favour the singular model and negative values the minimal model. With equal prior odds on the two models, $B_{\rm singular,minimal}$ is the posterior odds; $|\ln B|\sim1$ corresponds to odds of about three to one and is generally regarded as inconclusive, whereas $|\ln B|\gtrsim5$ corresponds to odds above about $150$ to one and is regarded as strong \citep{Trotta:2008qt}. We use $\ln B_{s,m}$ as a shorthand for \eqref{eq:bayes_factor}.

The evidence rewards fit and penalizes prior volume that the data do not use. Writing $\ln Z_{\mathcal M}=\ln L_{\max,\mathcal M}+\ln O_{\mathcal M}$, where $L_{\max}$ is the maximum likelihood, defines the Occam factor $O_{\mathcal M}\le1$, which measures the fraction of the normalized prior volume that remains compatible with the data \citep{2003itil.book.....M}. A coordinate that the data constrain well contributes a small factor; a coordinate whose posterior is prior-dominated contributes a factor close to unity. The penalty is therefore set by the effective number of constrained parameters rather than by the nominal dimensionality, a point that matters for the crust-only comparisons below, in which several singular coordinates are prior-dominated.

The priors that we use are uniform in the logarithms of the spin-noise parameters over the finite ranges listed in Table~\ref{tab:priors}, which makes them normalizable and gives $\pi(\theta|\mathcal M)=1/V_{\mathcal M}$ inside the prior box of volume $V_{\mathcal M}$. The ranges were established through preliminary expansion tests so that the likelihood support is not artificially truncated, and were then held fixed across all realizations of an experiment. The singular prior differs between the singular and minimal experiments because the high-likelihood region of the singular model shifts when it is fitted to minimal-model data. For the crust-only minimal model, the identifiable coordinates are $(\tau,A_{\rm min})$ and the prior on $x_{\rm s}$ is an external, normalized prior that integrates to unity; it does not enter the crust-only evidence. To quantify how the results depend on the prior, we repeat the crust-only comparisons with singular prior boxes whose half-widths are reduced or enlarged by one decade about a fixed centre.

\begin{table}
    \centering
    \small
    \begingroup
    \renewcommand{\arraystretch}{1.00}
    \setlength{\tabcolsep}{4pt}
    \caption{Prior ranges used in the simulation experiments (Sections~\ref{sec:singular_simulations}--\ref{sec:minimal_simulations}). All priors are uniform in the logarithm of the parameter, except where noted. The singular prior is centred differently in the two experiments; the minimal and $x_{\rm s}$ priors are shared. The $x_{\rm s}$ prior enters only the full-state minimal fits; in the crust-only minimal posteriors, the displayed $x_{\rm s}$ column is drawn uniformly in $x_{\rm s}$ over the same range and is not informed by the data.}
    \begin{tabular}{@{}lcc@{}}
        \hline
        Parameter & Singular data & Minimal data \\
        \hline\hline
        \multicolumn{3}{@{}l}{\textit{Singular model}} \\
        $\log_{10}(\tau/{\rm day})$ & $[1.4, 3.7]$ & $[1.4, 3.7]$ \\
        $\log_{10}r$ & $[-3.0, 0.5]$ & $[-3.5, 0.5]$ \\
        $\log_{10}(Q_{\rm c}/{\rm rad^2\,day^{-3}})$ & $[-12.0, -8.0]$ & $[-13.0, -7.0]$ \\
        $\log_{10}(Q_{\rm s}/{\rm rad^2\,day^{-3}})$ & $[-10.5, -6.5]$ & $[-11.0, -5.5]$ \\
        \hline\hline
        \multicolumn{3}{@{}l}{\textit{Minimal model}} \\
        $\log_{10}(\tau/{\rm day})$ & $[1.4, 3.7]$ & $[1.4, 3.7]$ \\
        $\log_{10}(A_{\rm min}/{\rm rad^2\,day^{-2}})$ & $[-11.0, -3.0]$ & $[-10.5, -5.0]$ \\
        $x_{\rm s}$ & $[10^{-3}, 0.25]$ & $[10^{-3}, 0.25]$ \\
        \hline
    \end{tabular}
    \label{tab:priors}
    \endgroup
\end{table}

Two estimators of the evidence are used. For full-state data, whose posteriors are sharply peaked, we use the Laplace approximation about the maximum a posteriori (MAP) estimate $\theta^\ast$ 
\citep{2003itil.book.....M},
\begin{equation}
 \ln Z_{\mathcal M}\simeq\ln p(\by|\theta^\ast,\mathcal M)+\frac{d}{2}\ln2\pi-\frac12\ln|\boldsymbol A|-\ln V_{\mathcal M} \,,\label{eq:laplace}
\end{equation}
where $d$ is the number of spin-noise parameters and $\boldsymbol A=-\nabla\nabla\ln p(\by|\theta,\mathcal M)|_{\theta^\ast}$ is the Hessian of the negative log likelihood, evaluated by finite differences. The last term is the normalization of the bounded prior. The approximation is reliable only when the posterior is unimodal and lies well inside the prior box; we monitor the distance of $\theta^\ast$ to the prior boundaries and the conditioning of $\boldsymbol A$ and treat  \eqref{eq:laplace} as a diagnostic whenever either check fails.

For crust-only data, the posteriors are broad, often bounded by the prior along one or more directions, and the Laplace approximation is not adequate. We instead use the stepping-stone estimator \citep{Xie2011SteppingStone}, which writes the evidence as a telescoping product of ratios of normalizing constants of power posteriors $p(\by|\theta)^{\beta}\pi(\theta)$ along a ladder $0=\beta_0<\beta_1<\dots<\beta_K=1$,
\begin{equation}
 \ln Z_{\mathcal M}=\sum_{k=0}^{K-1}\ln\left\langle p(\by|\theta,\mathcal M)^{\beta_{k+1}-\beta_k}\right\rangle_{\beta_k} \,,\label{eq:stepping_stone}
\end{equation}
where $\langle\cdot\rangle_{\beta_k}$ denotes an average over samples from the power posterior at $\beta_k$. The $\beta_0=0$ term is an average over prior samples, which is well defined because the prior is normalized, and each subsequent term is estimated from an affine-invariant ensemble Markov chain Monte Carlo (MCMC) run \citep{emcee2013} warm-started from the previous rung. The standard error of $\ln Z$ is propagated from the sample variances of the individual terms. Ladder, chain lengths and diagnostics are recorded in Appendix~\ref{app:evidence}. Bounded Laplace values for the crust-only fits are retained only as a check on the local geometry of the posterior.

\subsection*{Superfluid phase prediction}
\label{subsec:conditioning}

Radio timing observes the crust, but both models predict the superfluid phase given the crust. Since the crust and superfluid phases are jointly Gaussian (with the block covariance of Section~\ref{sec:phase_moments}), the hidden component is predicted by Gaussian conditioning. At fixed spin-noise parameters $\theta$, and after removing the GLS timing trend $\bX\widehat{\bbeta}$ from the crust observations, the conditional mean and covariance of the superfluid phase at any set of times $t_\ast$ are
\begin{align}
 \boldsymbol\mu_{{\rm s}|{\rm c}}&=K_{\rm sc}(t_\ast,t)\,\bK^{-1}\boldsymbol r \,,\label{eq:cond_mean}\\
 \boldsymbol C_{{\rm s}|{\rm c}}&=K_{\rm ss}(t_\ast,t_\ast)-K_{\rm sc}(t_\ast,t)\,\bK^{-1}K_{\rm cs}(t,t_\ast) \,,\label{eq:cond_cov}
\end{align}
where $\bK=K_{\rm cc}(t,t)+\diag(\sigma_n^2)$ is the crust covariance of the likelihood, $\boldsymbol r$ is the GLS residual of \eqref{eq:gls}, and the orientation of the cross block matters, the first argument of $K_{\rm sc}$ being the superfluid time. The same expressions with $K_{\rm cc}$ in place of $K_{\rm sc}$ and $K_{\rm ss}$ give the smoothed crust phase and its uncertainty, and for full-state data the conditioning is on the stacked crust and superfluid observations with the $2N\times2N$ covariance. Parameter uncertainty is propagated by evaluating (\ref{eq:cond_mean}-\ref{eq:cond_cov}) at draws from the posterior and mixing the resulting Gaussians.
The predicted superfluid phase inherits an unknown timing trend of its own, so predictions, latent phases and observations are all displayed after the common projection $\boldsymbol P$.

In the singular model the prediction depends on the full set $(\tau,r,Q_{\rm c},Q_{\rm s})$ and its uncertainty grows with that of the parameters. In the minimal model the prediction is rigid. Since $K_{\rm sc}=-(x_{\rm c}/x_{\rm s})K_{\rm cc}$, equation~\eqref{eq:cond_mean} reduces to $-(x_{\rm c}/x_{\rm s})$ times the smoothed crust phase, and the conditional covariance is $(x_{\rm c}/x_{\rm s})^2$ times that of the crust smoother. The shape of the superfluid prediction is thus fixed by the crust data alone and its amplitude by $x_{\rm s}$, which the crust data do not constrain and which must be supplied externally. This is the sense in which the minimal model turns a measurement of $x_{\rm s}$ into a definite hidden-state prediction.

\section{Singular simulations}
\label{sec:singular_simulations}

\begin{figure*}
 \centering
 \includegraphics[width=0.475\textwidth]{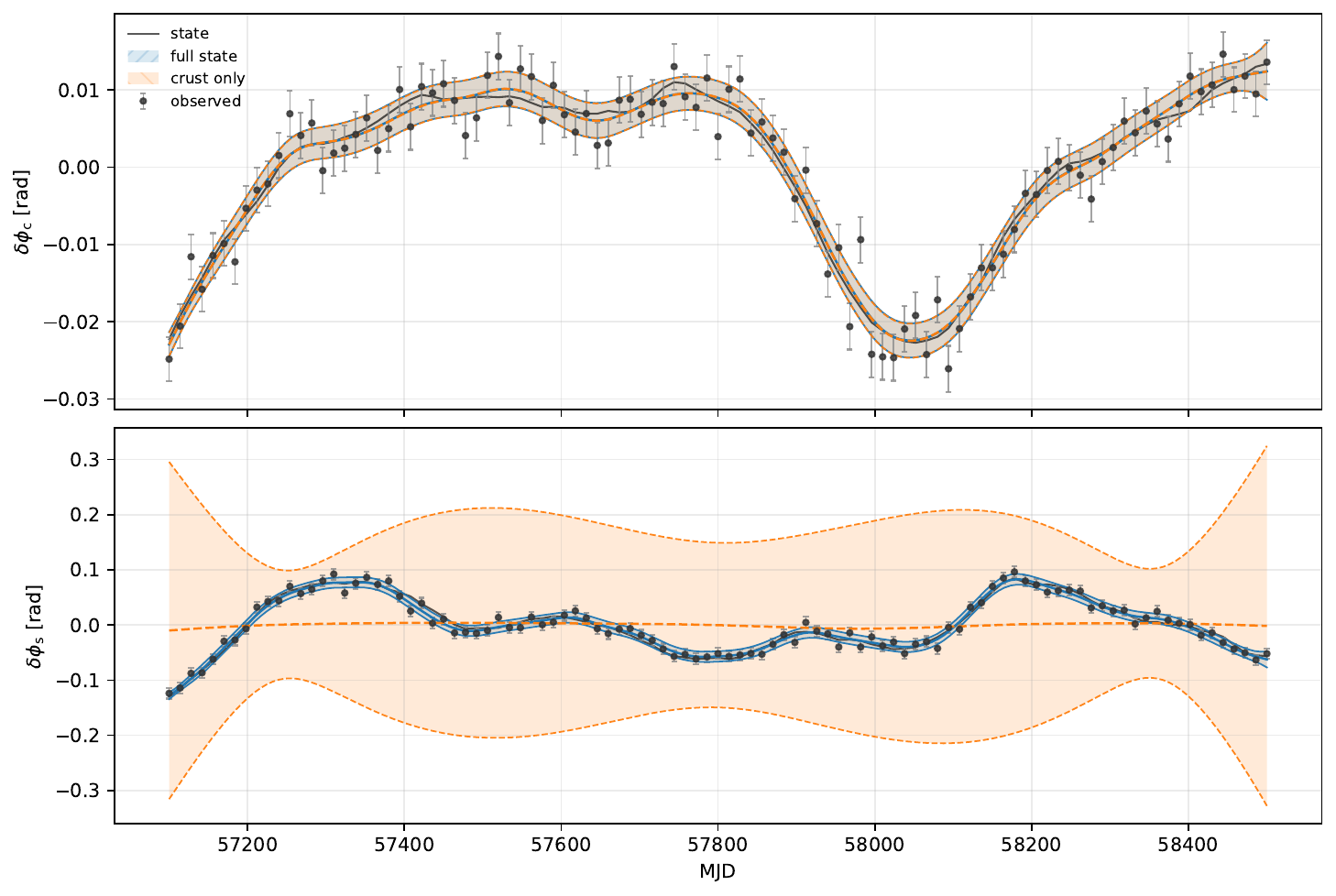}\hfill
 \includegraphics[width=0.475\textwidth]{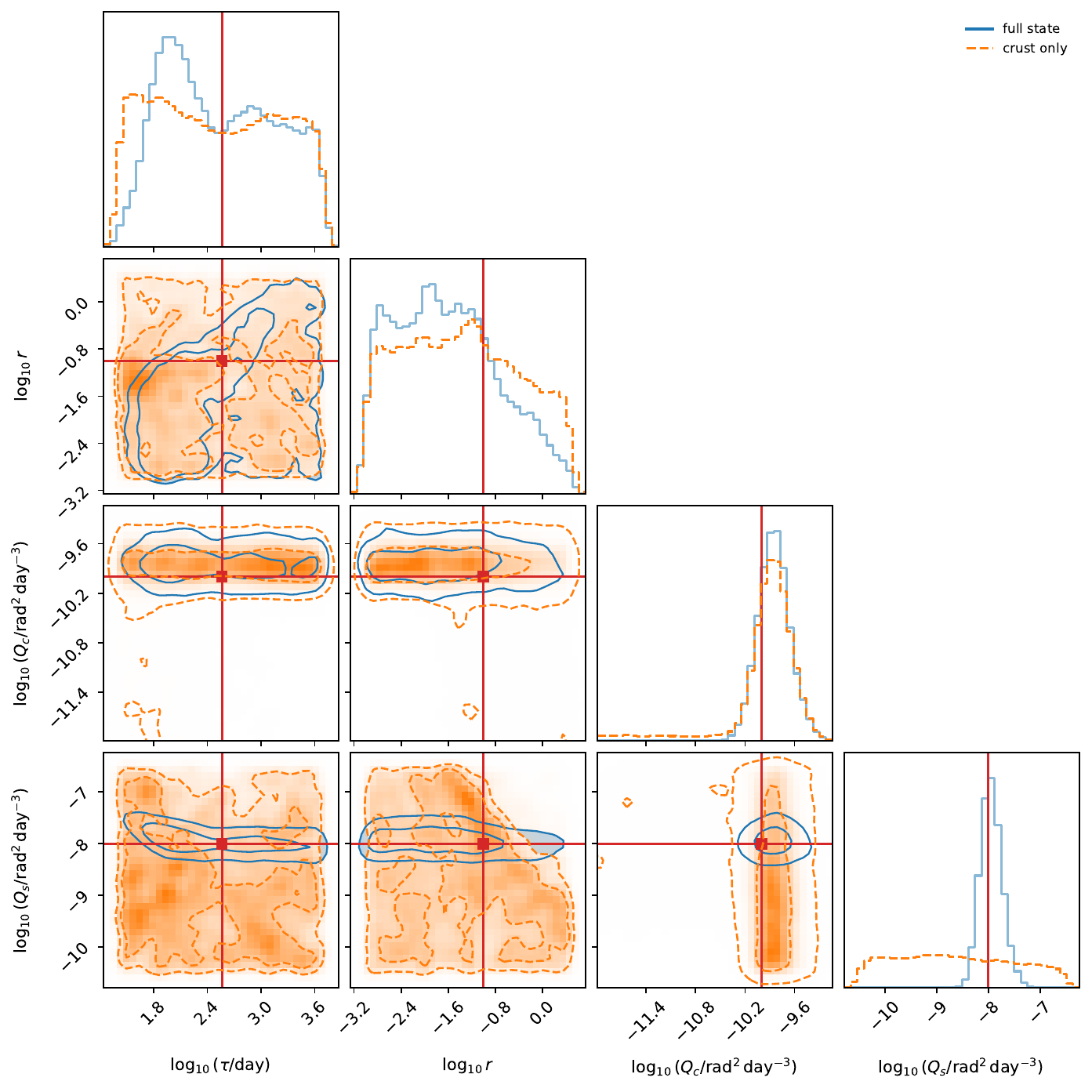}
 \caption{Singular inference with full-state and crust-only observations. [left] projected crust and superfluid phase reconstructions; for crust-only, the superfluid phase prediction is obtained by Gaussian conditioning. [right] posterior constraints for the same two observation settings; red markers give the injected coordinates.}
 \label{fig:singular_reconstruction_posterior}
\end{figure*}

\begin{figure*}
 \centering
 \includegraphics[width=0.95\textwidth]{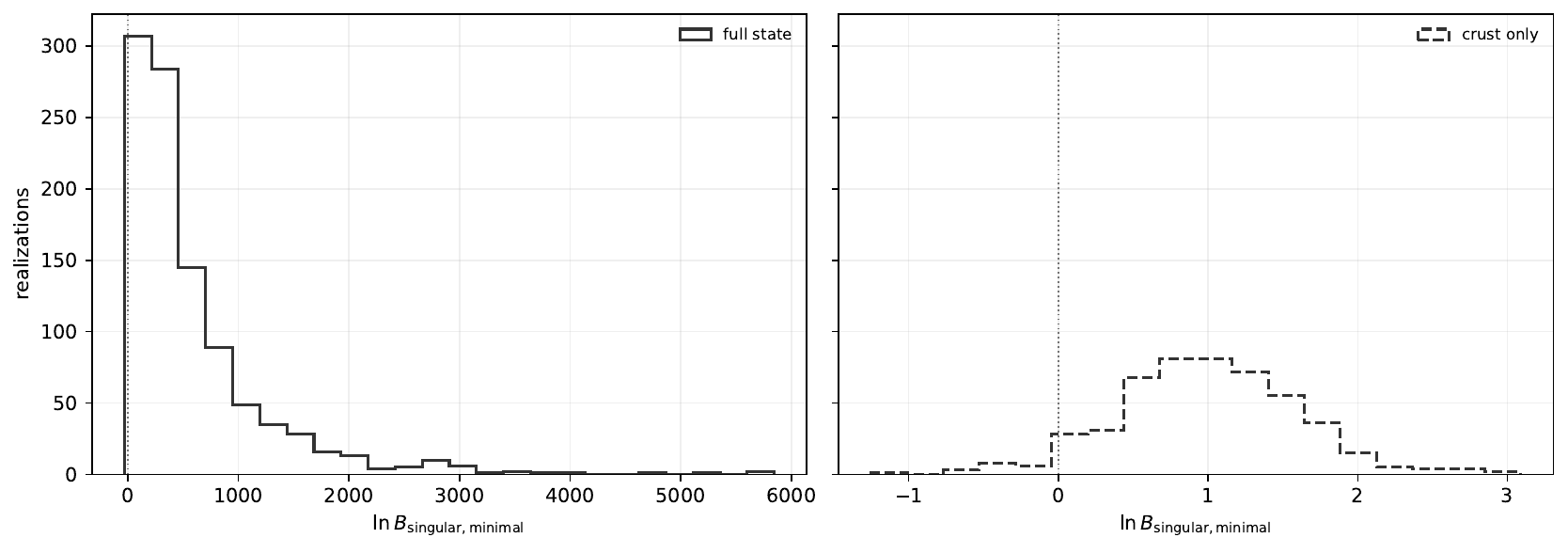}
 \caption{Singular phase simulations evidence ensembles. Positive $\ln B_{\rm singular,minimal}$ favours the singular model. The full-state panel contains 1000 bounded-Laplace calculations; the radio-only panel contains 500 stepping-stone calculations.}
 \label{fig:phase_evidence_singular}
\end{figure*}

Consider the singular model as the data-generating process. This section provides a controlled test of the phase-domain inference and reconstruction when the rotational state is fully observed (full-state), and of the systematic effects that arise when only the crust is observed (crust-only). The noisy singular realization shown in Figure~\ref{fig:paths}, with the model and noise parameters of Table~\ref{tab:simulation_parameters} and the cadence of Section~\ref{subsec:simulated_data}, is analyzed with the singular model in two observation settings: full state, in which both the crust and superfluid phases are observed, and crust only, in which the superfluid is hidden and must be predicted. In both settings the quadratic timing model of each observed component is marginalized analytically, so that the inference concerns the spin-wandering parameters $(\tau, r, Q_{\rm c}, Q_{\rm s})$ alone. The same fitted objects are used for the reconstructions, parameter posteriors and evidence calculations.

Figure~\ref{fig:singular_reconstruction_posterior} shows the phase reconstructions and the corresponding full-state and crust-only posterior constraints. With full-state observations, the smoothed crust and superfluid phases follow the latent phases within bands that are comparable to the measurement uncertainties; the conditioning of Section~\ref{subsec:conditioning} then acts as a smoother for both components. With crust-only observations, the crust phase is reconstructed equally well, since the crust data are the same and the crust covariance is well described by either model. The superfluid prediction, however, is of a different nature. Its conditional mean stays close to zero and its band is wide, spanning several times the excursions of the latent superfluid phase. This is the conditional covariance of \eqref{eq:cond_cov} at work. The cross covariance $K_{\rm sc}$ transfers information from the crust to the superfluid only through the exchange mode, which is subdominant in the crust phase for the parameters considered here, so that most of the superfluid variance remains unexplained by the crust data. The width of the band is therefore a faithful statement of what the crust alone can say about the superfluid in this model.

The parameter posteriors reflect the same hierarchy. With full-state observations, the two diffusion coefficients are determined to within a factor of about two, with the superfluid coefficient $Q_{\rm s}$ pinned by the superfluid channel. The relaxation time and the ratio $r$ remain broad even in this setting, with posteriors that lean toward shorter $\tau$ and smaller $r$ than the injected values; for a window of about four relaxation times and a diffusive mode that dominates the low-frequency phase, the exchange mode that carries $\tau$ and $r$ leaves a comparatively weak imprint. With crust-only observations, the crust diffusion coefficient $Q_{\rm c}$ is constrained about as well as with full-state data, because it sets the amplitude of the crust phase directly. In contrast, the superfluid coefficient becomes prior-dominated. Its marginal is nearly flat over the prior range, and its two-dimensional marginals with $\tau$ and $r$ show that the crust phase constrains only the combination $\tau_{\rm s}^2Q_{\rm s}$ that enters the transformed intensities $Q_\pm$ and $Q_\times$. The relaxation time and $r$ are likewise prior-dominated. The injected parameters lie inside the 95 percent intervals in both settings, so the crust-only posteriors are wide rather than biased; what is lost with the superfluid is the ability to attribute the observed crust wander to a specific partition of the stochastic torques between the components.

The same realization is also fitted with the minimal model, and Figure~\ref{fig:phase_evidence_singular} presents the log Bayes factor of equation~\eqref{eq:bayes_factor} over independent realizations of the singular sampled mock data, with independent measurement-noise draws, using the estimators of Section~\ref{subsec:evidence}. With full-state observations the Bayes factor is overwhelming. The median $\ln B_{s,m}$ is of order several hundred and essentially all realizations favour the singular model. The minimal model has no diffusive common mode and cannot reproduce the joint crust-superfluid covariance of a singular realization, so that the full-state data rule it out decisively. With crust-only observations the picture changes qualitatively. The median $\ln B_{s,m}$ is of order unity, corresponding to posterior odds of only about three to one in favour of the singular model, even though the sign of the Bayes factor selects the singular model in about 96 percent of the realizations. The consistency of the sign across realizations indicates that the crust phase does carry a signature of the diffusive mode, but the magnitude of the Bayes factor indicates that this signature is weak in any single realization of the length and cadence considered here.

The weak crust-only preference deserves comment, because both models reconstruct the crust phase equally well and one might expect the Occam factor to penalize the singular model for its two additional parameters. The penalty does not materialize, for the reason anticipated in Section~\ref{subsec:evidence}. The crust-only singular posterior is prior-dominated along $\tau$, $r$ and $Q_{\rm s}$, and a coordinate whose normalized prior is returned essentially unchanged contributes an Occam factor close to unity. The effective number of constrained singular parameters is therefore closer to that of the minimal model than the nominal count suggests, and the Bayes factor is left to be decided by the modest difference in the maximum likelihoods and by the residual prior-volume differences of the partially constrained coordinates. The same reasoning implies that the crust-only Bayes factor must depend on the prior ranges, and we quantify this directly. Reducing the half-widths of the singular prior by one decade about its centre raises the median crust-only $\ln B_{s,m}$ by about a factor of two and increases the fraction of realizations favouring the singular model to unity; enlarging the half-widths by one decade lowers the median by a similar factor and reduces the fraction to about 88 percent. The sign of the median is unchanged across the tested ranges, but its magnitude is not, and both are reported together as the crust-only result. A crust-only preference for the singular model of the strength found here should accordingly be read as prior-limited evidence rather than as a detection of a diffusive common mode.

\section{Minimal simulations}
\label{sec:minimal_simulations}

\begin{figure*}
 \centering
 \includegraphics[width=0.49\textwidth]{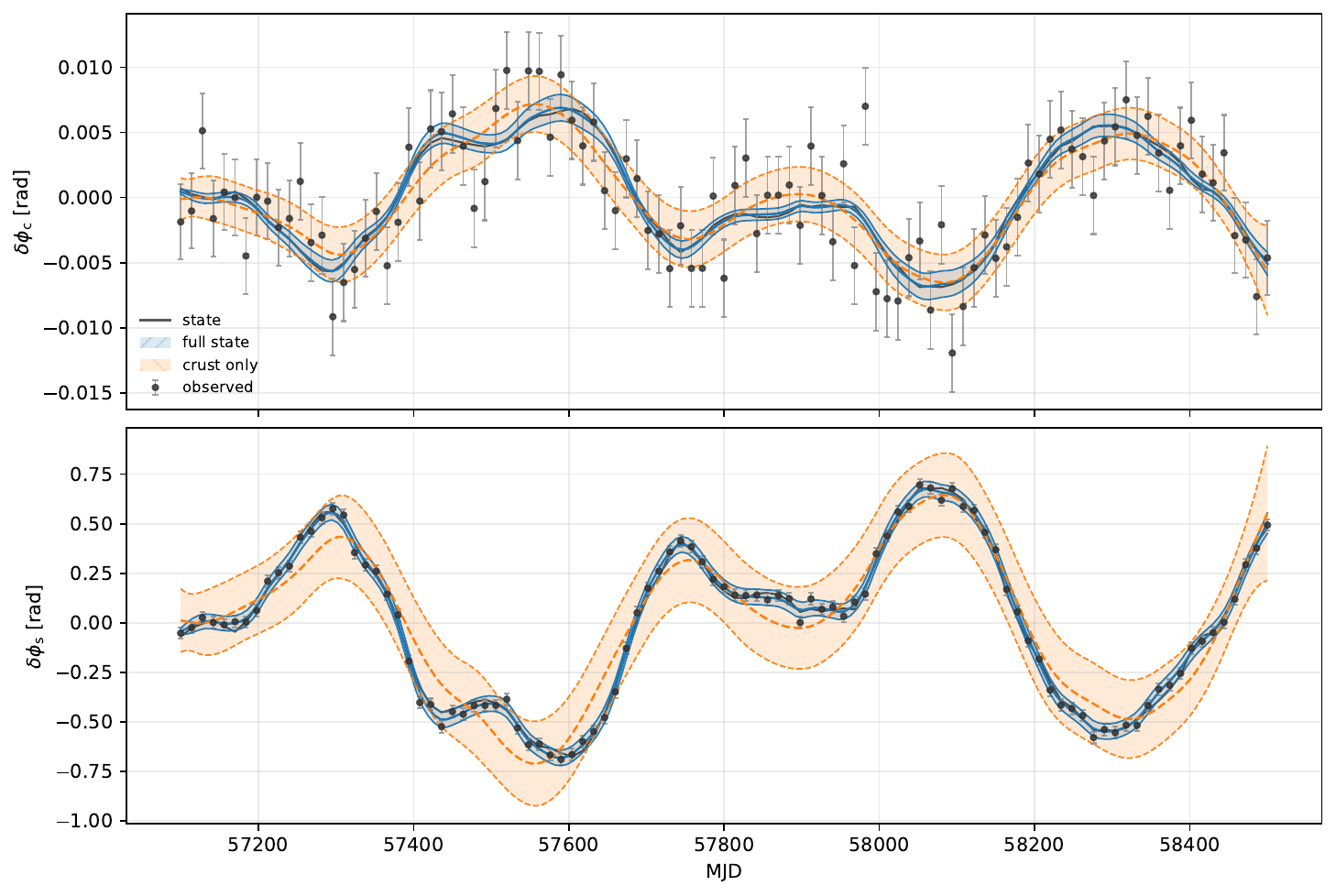}\hfill
 \includegraphics[width=0.49\textwidth]{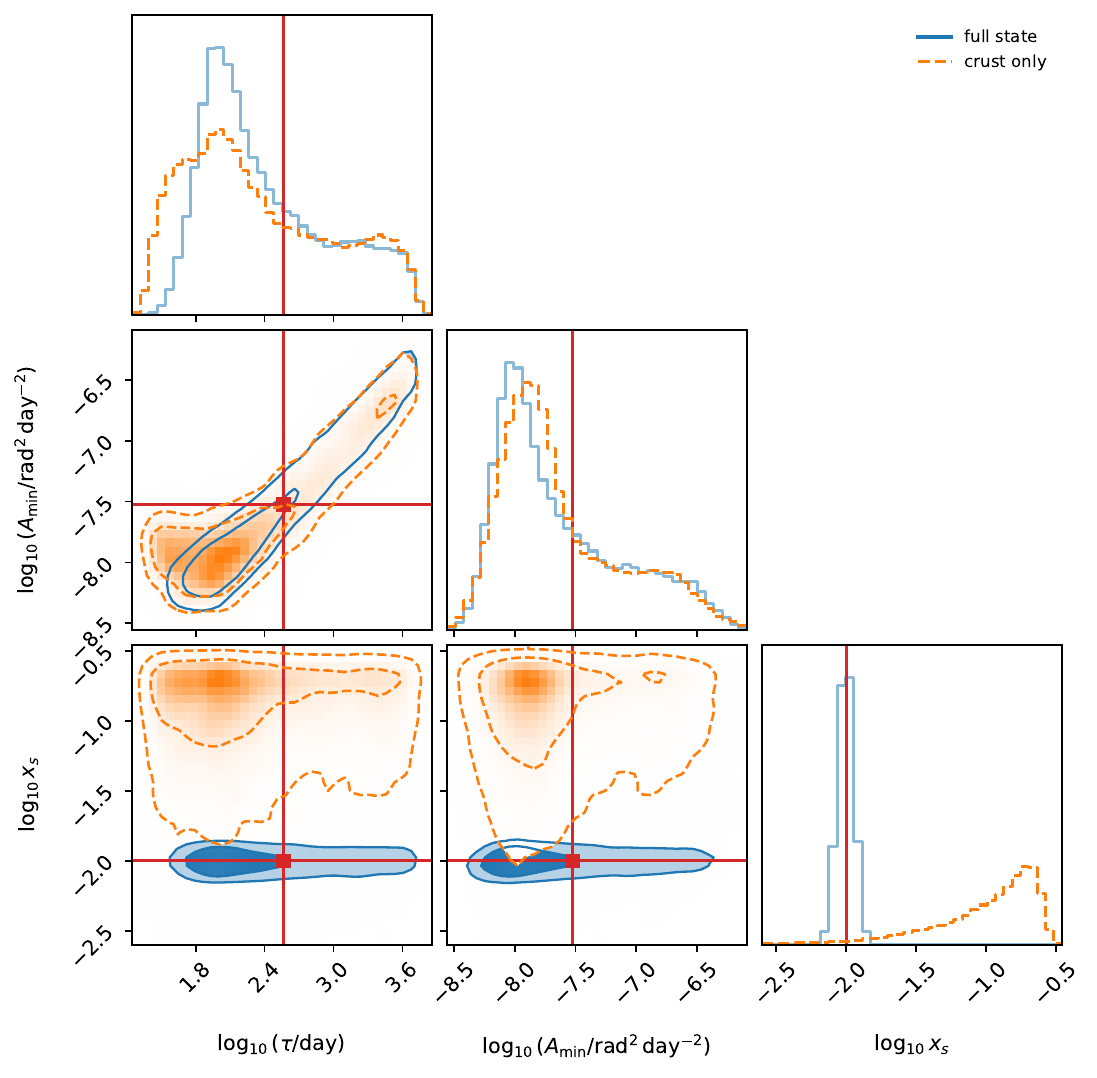}
 \caption{Minimal model inference with full-state and crust-only observations. [left] projected crust and superfluid phase reconstructions; for crust-only, the superfluid phase prediction is obtained by Gaussian conditioning with $x_{\rm s}$ fixed at its injected value. [right] posterior constraints for the same two observation settings; red markers give the injected coordinates. The crust-only $x_{\rm s}$ column is a draw from its prior, since the crust phase covariance depends on $(\tau, A_{\rm min})$ only.
 }
 \label{fig:minimal_reconstruction_posterior}
\end{figure*}

\begin{figure*}
 \centering
 \includegraphics[width=0.95\textwidth]{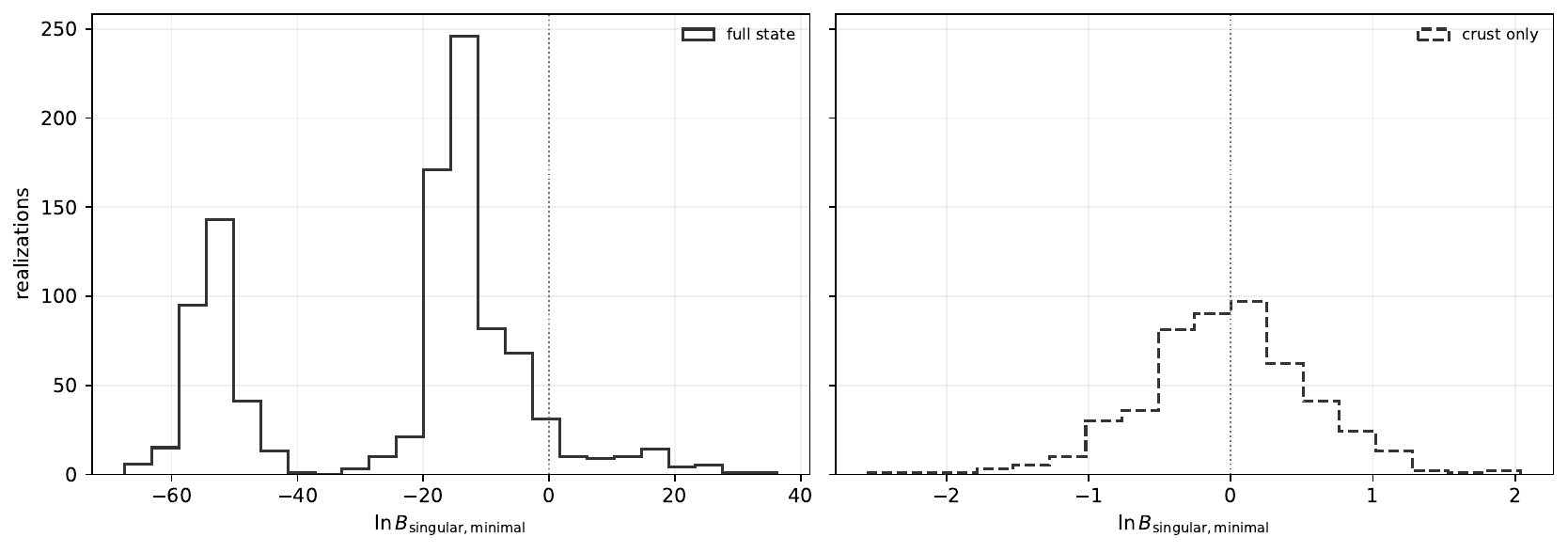}
 \caption{Minimal phase simulations evidence ensembles. Positive $\ln B_{\rm singular,minimal}$ favours the singular model, negative the minimal model. The full-state panel contains 1000 bounded-Laplace calculations; the radio-only panel contains 500 stepping-stone calculations.}
 \label{fig:phase_evidence_minimal}
\end{figure*}

We now take the minimal model as the data-generating process and repeat the experiment of Section~\ref{sec:singular_simulations}. The noisy minimal realization of Figure~\ref{fig:paths}, generated with the parameters of Table~\ref{tab:simulation_parameters}, is analyzed with the minimal model in the full-state and crust-only settings. The angular velocity exchange mode of this realization is stationary, but its integrated phase still wanders, and the removal of the three timing modes changes its appearance substantially. The projected crust phase fluctuates on the scale of the relaxation time with an amplitude only a few times the measurement uncertainty, whereas the superfluid phase, which is $x_{\rm c}/x_{\rm s}$ times larger in the minimal model, is measured with a much higher signal-to-noise ratio. In the full-state setting the inference concerns $(\tau, A_{\rm min}, x_{\rm s})$; in the crust-only setting, only $(\tau, A_{\rm min})$ enter the likelihood, and $x_{\rm s}$ is carried along as an external prior draw for display and for the superfluid prediction, as explained in Section~\ref{subsec:conditioning}.

Figure~\ref{fig:minimal_reconstruction_posterior} shows the reconstructions and posteriors. With full-state observations, the smoothed crust and superfluid phases follow the latent phases closely, and the superfluid band is narrow because the superfluid channel is measured with a high signal-to-noise ratio. With crust-only observations, the crust is reconstructed with a band comparable to the crust measurement uncertainty, as in the singular case. The crust-only superfluid prediction is qualitatively different from the singular one; rather than a broad band around zero, it is a rescaled copy of the smoothed crust phase, with the amplitude set by the supplied value of $x_{\rm s}$, here the injected value, and with a band that is the rescaled crust band. When the supplied $x_{\rm s}$ matches the generating value, as in the figure, the prediction follows the latent superfluid phase; a different $x_{\rm s}$ would rescale the prediction without changing its shape. This rigidity is what makes the minimal model falsifiable with an external handle on the hidden state.

The parameter posteriors show a strong degeneracy between $\tau$ and $A_{\rm min}$ in both settings, with a ridge along $A_{\rm min}\propto\tau$ that runs through the injected values, with the bulk of the posterior at shorter relaxation times and smaller amplitudes and a tail extending to the upper end of the $\tau$ prior. The origin of the ridge is that the observation window covers only a few relaxation times. An OU process observed over a span not much longer than $\tau$ is difficult to separate from a random walk with diffusion coefficient $A_{\rm min}/\tau$, and the phase data constrain this ratio far better than $\tau$ and $A_{\rm min}$ individually. The marginal posteriors of $\tau$ and $A_{\rm min}$ are nearly identical in the full-state and crust-only settings, because the superfluid channel is an exact rescaling of the crust in the minimal model and carries no independent information about the lag dynamics. What the superfluid channel does supply is the rescaling factor itself. With full-state data, $x_{\rm s}$ is determined to a few percent, since the ratio of the two observed phases is $-x_{\rm c}/x_{\rm s}$ epoch by epoch. With crust-only data the displayed $x_{\rm s}$ marginal reproduces its prior, as it must, and its two-dimensional marginals with $\tau$ and $A_{\rm min}$ are uncorrelated. The injected parameters lie inside the 95 percent intervals in both settings.
The direction of the ridge has a simple physical reading through the definition \eqref{eq:Amin}. At fixed $Q$ and $x_{\rm c}$, the amplitude $A_{\rm min}=Q\tau/x_{\rm c}^2$ grows in proportion to the relaxation time, because a lag that is driven by the same stochastic torque but damped more slowly accumulates a larger stationary variance. Moving along $A_{\rm min}\propto\tau$ is therefore moving at constant $Q/x_{\rm c}^2$, that is, at a constant strength of the internal stochastic torque as seen by the crust. Over an observation span of a few relaxation times, the phase data measure this torque strength, which sets the short-time diffusion of the lag, considerably better than they measure the damping that eventually bounds it. The degeneracy is thus between the torque and the damping, and a longer baseline relative to $\tau$ is what would break it.

The same realization is also fitted with the singular model, and Figure~\ref{fig:phase_evidence_minimal} presents the log Bayes factor over independent minimal realizations. With full-state observations, the Bayes factor favours the minimal model decisively, with a median $\ln B_{s,m}$ of order minus ten to minus twenty and the minimal model selected in the large majority of realizations. The singular model can reproduce a single minimal realization closely, by driving its common-mode diffusion toward small values, but it pays for the additional freedom through the Occam factor and through the mismatch of its joint crust--superfluid covariance with the rank-one structure of the minimal process. The magnitude of the full-state Bayes factor is smaller than in the singular experiment, which reflects the asymmetry of nested comparisons. A singular realization cannot be reproduced by the minimal model at all, whereas a minimal realization can be approximated by the singular model.

With crust-only observations, the comparison is uninformative. The median $\ln B_{s,m}$ is consistent with zero and the fraction of realizations favouring the minimal model is consistent with one half. The reason is structural rather than statistical. The crust covariance of the minimal model depends on $(\tau, A_{\rm min})$ only, and the singular crust covariance contains, over a substantial part of its prior volume, realizations whose exchange-mode contribution reproduces the minimal crust covariance to within the measurement uncertainty while the diffusive contribution stays below detectability. The observed crust process can thus be described equally well by both models, the maximum likelihoods are nearly equal, and the Occam factor of the singular model is close to unity along its prior-dominated directions. The models differ in their predictions for the hidden superfluid, which the crust data do not probe. This is the counterpart, from the opposite direction, of the singular experiment. Singular data leave a weak but consistent common-mode imprint in the crust, whereas minimal data leave nothing in the crust that the singular model cannot also produce.

The prior study is correspondingly revealing. Reducing the singular prior half-widths by one decade shifts the median crust-only $\ln B_{s,m}$ to positive values, so that the majority of realizations favour the singular model, whereas enlarging them by one decade shifts the median to negative values and the majority to the minimal model; the nominal prior sits in between with no preference. The sign of the crust-only comparison is therefore set by the prior volume assigned to the singular model rather than by the data, precisely because the data do not constrain the coordinates that distinguish the two physical models. Taken together with the singular experiment (Section \ref{sec:singular_simulations}), this establishes the central result of the simulations. Full-state observations recover the parameters of the generating process, reconstruct both phases and distinguish the stationary and nonstationary spin noise models reliably, whereas crust-only phase observations reconstruct the crust and, at most, leave a weak and prior-limited preference when the generating process is singular.

\section{Phase or angular velocity}
\label{sec:phase_or_angular_velocity}
\label{sec:information_loss}

The physical distinction between the two models is a statement about angular velocities. The minimal spin noise is stationary and the singular spin noise is not. Radio timing, however, delivers phases, or equivalently timing residuals. There are then two routes to a test of the models. The phase can be analyzed directly, as with \cite{Dong:2025nho} using the Kalman filter or with the likelihood of Section~\ref{sec:likelihood}, or it can first be processed into a time series of local spin-frequency estimates, as was done for PSR J1359-6038 by \citet{ONeill:2024uiw}.
The question we address in this section is whether the features that distinguish the models, stationarity or its absence, survive the chain of operations that separates the physical state from the analyzed data, and whether the two routes lead to the same conclusions. We do so by processing the same simulated phase data along both routes and comparing the results.

It is useful to regard each step of the chain as a linear operator acting on the two-component state, since every such operator maps a Gaussian process onto another Gaussian process with a covariance that follows exactly. If $\boldsymbol O$ is linear and $\bK$ is the covariance of its input, the output covariance is $\boldsymbol O\bK\boldsymbol O^{\mathsf T}$. The data analyzed along the phase route are $\boldsymbol P\boldsymbol S\,\delta\boldsymbol\phi$, with $\boldsymbol S$ the selection of the crust and $\boldsymbol P$ the timing projection; along the angular velocity route they are $\boldsymbol H\boldsymbol S\,\delta\boldsymbol\phi$, with $\boldsymbol H$ a differentiation operator, followed by the removal of a linear trend. We discuss the operators in turn.

\subsection*{Selection of the crust}

Hiding the superfluid maps the multi-output covariance of Section~\ref{sec:phase_moments} onto its crust block $K_{\rm cc}$. This is an observability loss of physical origin and it is the same along both routes. It removes the parameter combinations that appear only in the cross and superfluid blocks. In the singular model, $\boldsymbol S$ removes the partition of the stochastic torques between the components, which is why $Q_{\rm s}$ and $r$ become prior-dominated in Section~\ref{sec:singular_simulations}. In the minimal model, $\boldsymbol S$ isolates the superfluid moment of inertia fraction $x_{\rm s}$, which is absent from $K_{\rm cc}$ altogether. Sections~\ref{sec:singular_simulations}-\ref{sec:minimal_simulations} have shown that this operation alone reduces a decisive model comparison to a weak or absent one.

\subsection*{Integration to phase}

The phase is the time integral of the angular velocity, so the phase covariance is the double integral of the angular velocity covariance, and conversely $\partial^2K_\phi/\partial t\,\partial t'=K_\Omega$ (Appendix~\ref{app:verification}). For a continuously observed process, integration is invertible and no information is lost. What changes is the distribution of that information. In the frequency domain integration multiplies the spectrum by $(2\pi f)^{-2}$, so that the phase is dominated by the lowest frequencies the observation window admits; in the time domain it converts a stationary angular velocity into a phase whose variance grows with time, as \eqref{eq:Kcc} and \eqref{eq:H} make explicit, and it introduces a dependence on the start of the observations. The likelihood leverage of the phase data thus sits at low frequency, which is precisely where the timing model acts.

\subsection*{Timing projection}

A pulsar timing model fits at least a phase offset, a spin frequency and a spin-down, and only the residual is available for spin-noise analysis. Section~\ref{sec:likelihood} treats this exactly by marginalizing the three coefficients, and the OLS projector $\boldsymbol P$ displays its effect. This process also produces nonstationary residuals \citep{Lee:2012bf, vanHaasteren:2012hj, Allen:2025waa}; but of nonstationarity that is dinstinguishable compared to the ones in the spin frequencies that we are interested at. The projection removes three modes from the phase, but these are the modes on which a red process has the greatest leverage, and the transmission of the projected data falls as $f^6$ below the inverse observation span \citep{Hazboun:2019vhv, Pitrou:2024scp, Allen:2025waa}. A spin-frequency time series, by contrast, needs only a frequency offset and a spin-down, so its trend has one fewer degree of freedom and the corresponding transmission falls as $f^4$ \citep{Pitrou:2024scp}. The timing projection therefore removes more low-frequency information from the phase route than the frequency-trend removal does from the angular velocity route; but the angular velocity route has already lost information in the differentiation step, to which we now turn.

\subsection*{Differentiation}

The simplest estimate of the angular velocity from sampled phases is the finite difference of adjacent epochs,
\begin{equation}
 \widehat{\delta\Omega}_{n+1/2}=\frac{y_{n+1}-y_n}{t_{n+1}-t_n} \,,
\end{equation}
which has two well-known drawbacks. Its measurement uncertainty is $\sqrt{\sigma_{n+1}^2+\sigma_n^2}/(t_{n+1}-t_n)$, which for a cadence of 14 days and the phase uncertainties of Table~\ref{tab:simulation_parameters} is comparable to the angular velocity excursions themselves, and adjacent differences share one phase datum and are therefore anticorrelated, with correlation coefficient $-1/2$ for equal uncertainties. Treating the finite differences as independent measurements discards this correlation and miscalibrates the likelihood. The standard remedy is to fit a straight line to the phases within a window of several epochs and to take its slope as the local spin frequency at the centre of the window \citep{ONeill:2024uiw}. Each slope is a weighted least-squares estimate, and it is a fixed linear combination of the phases in its window; collecting these combinations as rows gives the operator $\boldsymbol H$ with as many rows as there are windows. Here we use non-overlapping windows of at least 20 days containing at least five phase epochs, weight the phases by their inverse variances and assign each slope to the weighted mean epoch of its window. With this choice the 101 phase epochs of the simulations are compressed into 20 local spin-frequency estimates with an effective cadence of about 70 days. Since the windows do not overlap, the slopes have independent measurement errors, with variances given by the diagonal of $\boldsymbol H\diag(\sigma_n^2)\boldsymbol H^{\mathsf T}$; the price of this independence is the compression of the cadence.

Differentiation acts on the spectrum as $(2\pi f)^2$ and thus undoes the integration. But, a window slope is not an instantaneous angular velocity, but rather the average of the angular velocity over the window, weighted by a linear kernel. Window averaging is a low-pass operation that suppresses fluctuations on time scales shorter than the window, and the compression of the cadence removes the ability to resolve them. For the relaxation time of one year used here the windows are short compared with $\tau$, and the covariance of the slopes, $\boldsymbol H\bK_\phi\boldsymbol H^{\mathsf T}$, differs from the analytic covariance of the instantaneous angular velocity at the window epochs by only a few percent.
The angular velocity route of this section nonetheless treats the slopes as instantaneous angular velocities at their epochs, with independent measurement errors and a linear trend, so that it corresponds to how such processed data would be analyzed in practice; the exact covariance $\boldsymbol H\bK_\phi\boldsymbol H^{\mathsf T}$ is retained as a diagnostic of the approximation.

\subsection*{Paired experiment}

\begin{figure*}
 \centering
 \includegraphics[width=0.95\textwidth]{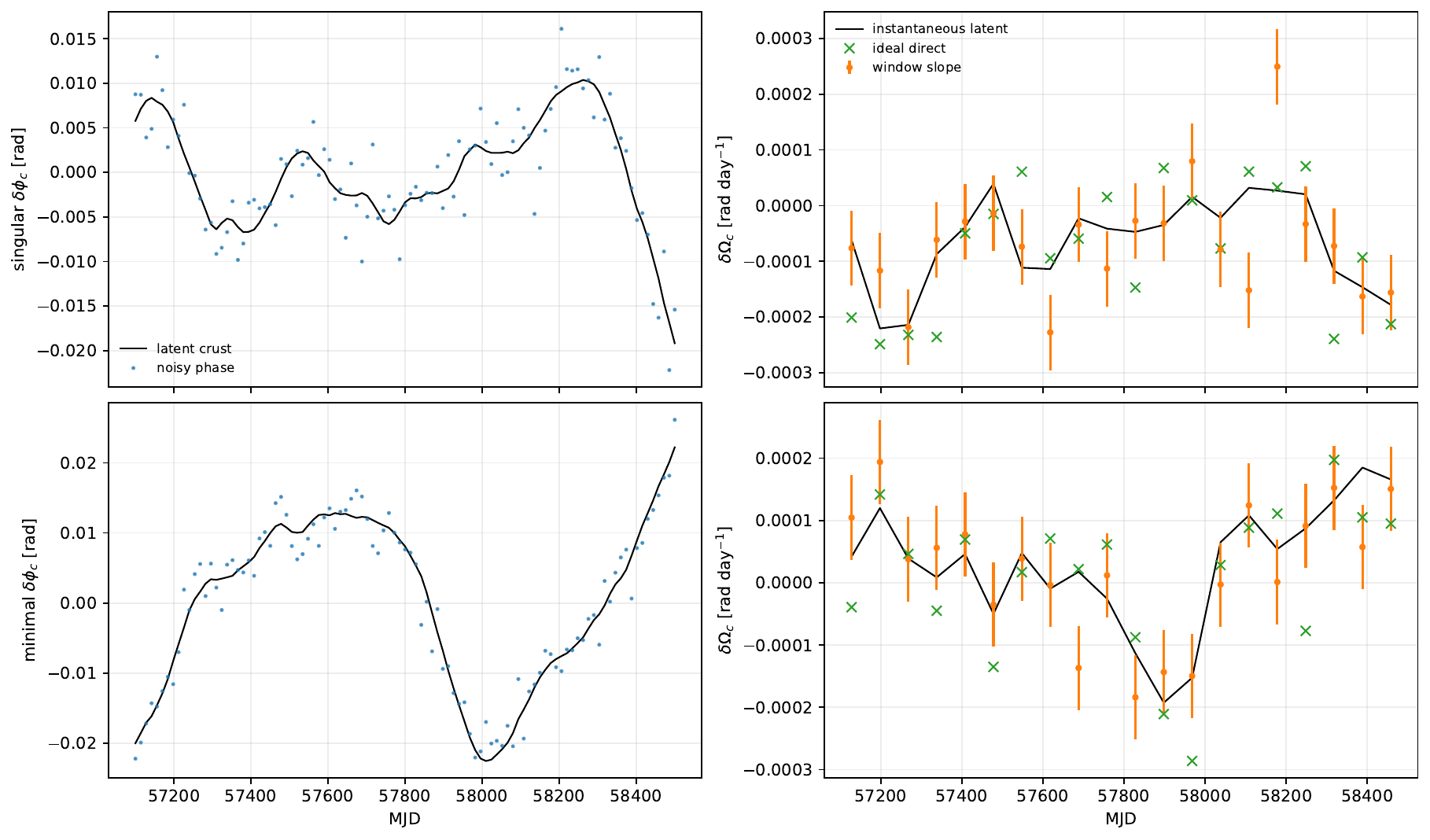}
 \caption{
Paired phase-angular velocity experiment for singular (top) and minimal (bottom) realizations. [left] projected crust phase observations and latent crust phase. [right] window-derived local spin-frequency estimates with their errors, the instantaneous latent angular velocity at the window epochs, and an ideal control that observes the instantaneous angular velocity at the same epochs with the same nominal errors. The slopes are derived from the phases on the left; both panels share the latent realization and the measurement-noise draw.
 }
 \label{fig:phase_velocity_operator}
\end{figure*}

\begin{figure*}
 \centering
 \includegraphics[width=0.49\textwidth]{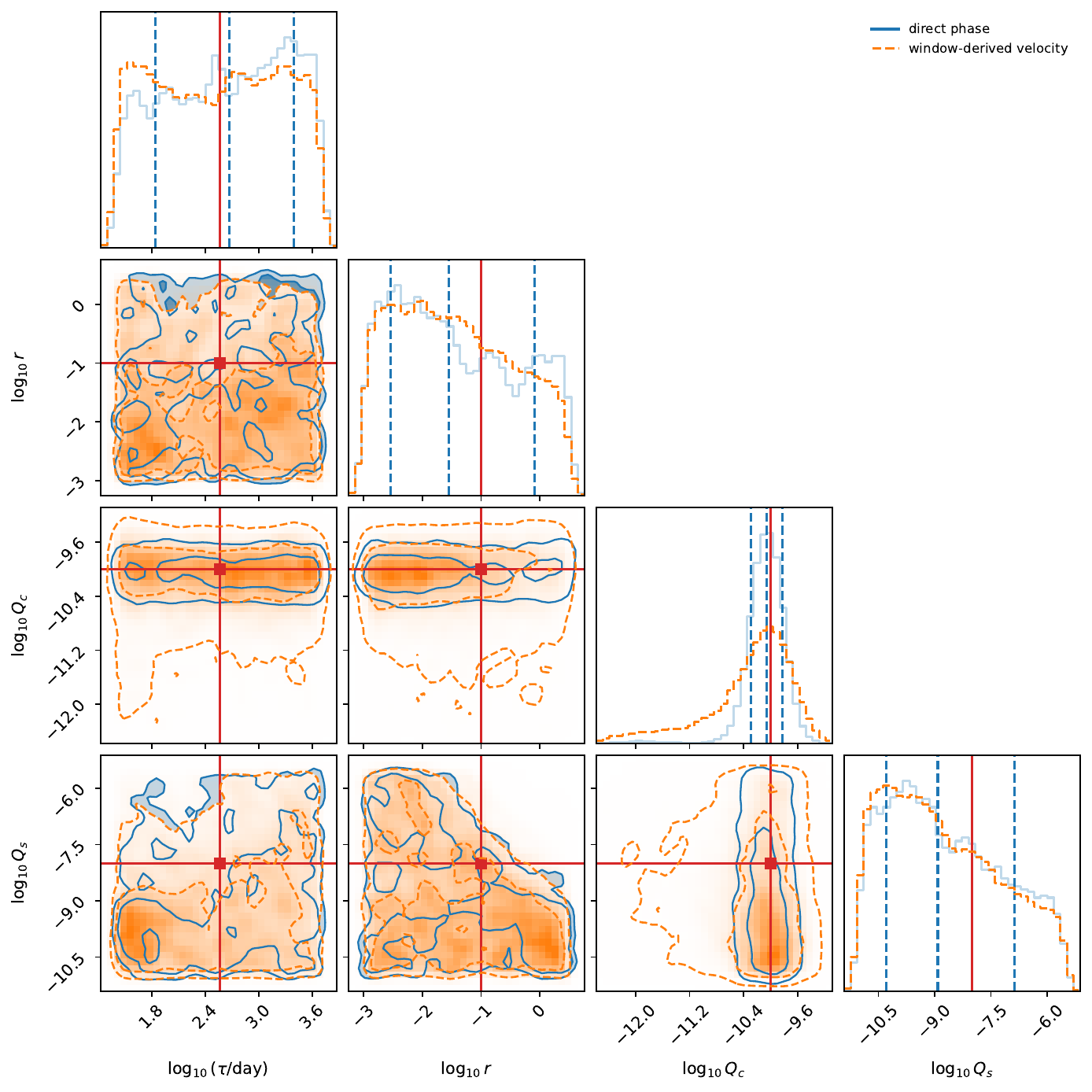}\hfill
 \includegraphics[width=0.49\textwidth]{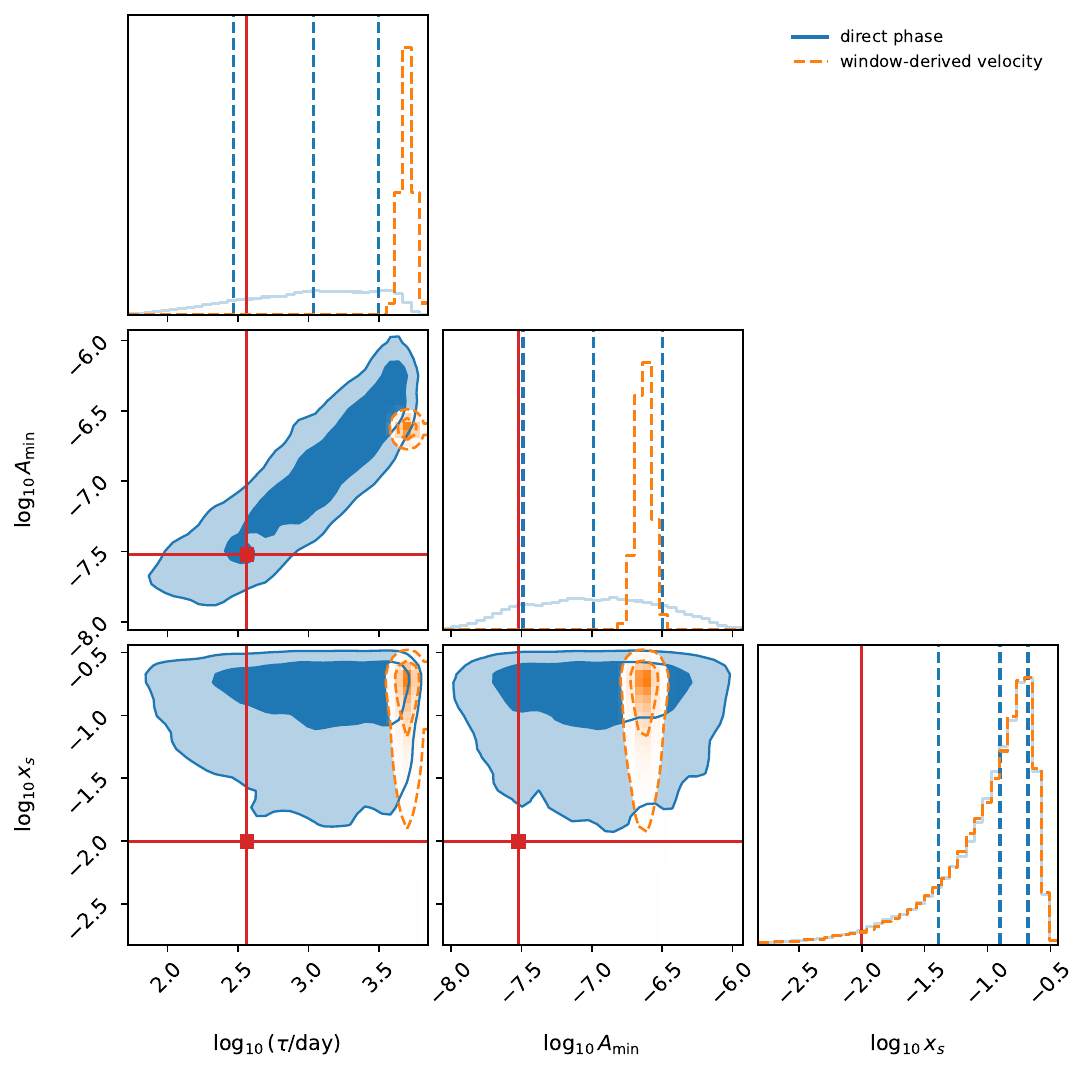}
 \caption{Crust-only posteriors from direct phase (solid blue) and window-derived angular velocity (dashed orange) for singular (left) and minimal sampled mock data (right). Each realization is analyzed with its generating model, and red markers give the injected coordinates.
 }
 \label{fig:phase_velocity_posteriors}
\end{figure*}

We generate joint realizations of the latent phase and angular velocity from their exact cross covariance, add measurement noise to the phase only, and analyse the same realization along the two routes/domains: directly in phase, with the likelihood of Section~\ref{sec:likelihood} and a quadratic timing model, and in window-derived angular velocity, with the corresponding analytic angular velocity likelihood \citep{Bernardo:2026opv} and a linear trend. Figure~\ref{fig:phase_velocity_operator} shows one such realization for each generating model, with the projected crust phase beside the window slopes and the instantaneous angular velocity at the window epochs. The two routes share the latent realization and the measurement-noise draw, so that any difference between them is due to the operators and not to sampling. Table~\ref{tab:evidence_summary} collects the Bayes factors \eqref{eq:bayes_factor} over independent realizations for both routes, both generating models and both observation settings. The phase rows repeat the results of Sections~\ref{sec:singular_simulations}-\ref{sec:minimal_simulations}.

\begin{table}
\centering
\caption{
Median log Bayes factors over independent realizations for the phase and window-derived angular velocity routes, for singular and minimal generating models and for full-state and crust-only observations. The column `recovery' is the fraction of realizations in which the sign of $\ln B_{s,m}$ selects the generating model. Full-state values use bounded Laplace evidences; crust-only values use stepping-stone sampling.
}
\label{tab:evidence_summary}
\begin{tabular}{lllrrr}
\toprule
domain & generator & setting & runs & $\ln B_{s,m}$ & recovery \\
\midrule
phase & singular & full & 1000 & 368.513 & 99.4\\
phase & singular & crust & 500 & 1.002 & 95.8\\
phase & minimal & full & 1000 & -15.882 & 93.8\\
phase & minimal & crust & 500 & -0.027 & 51.6\\
velocity & singular & full & 1000 & 39.739 & 94.5\\
velocity & singular & crust & 500 & 1.015 & 88.8\\
velocity & minimal & full & 1000 & -45.452 & 99.2\\
velocity & minimal & crust & 500 & 0.003 & 49.8\\
\bottomrule
\end{tabular}
\end{table}

The table shows that the qualitative conclusions of Sections~\ref{sec:singular_simulations}-\ref{sec:minimal_simulations} survive the passage to window-derived angular velocities. With full-state data, both routes prefer the generating model in the large majority of realizations, with Bayes factors that are decisive in magnitude. With crust-only data, both routes are weak or uninformative. A singular generator leaves a consistent but small preference in its favour, somewhat less consistent along the angular velocity route, and a minimal generator leaves no preference along either route. The stationarity or nonstationarity of the underlying spin noise is therefore neither created nor destroyed by integration, projection and differentiation. What these operations change is how much of the distinction is retained in a finite, noisy data set. Within the full-state rows, the phase route gives a much larger singular-generator Bayes factor and a smaller minimal-generator Bayes factor than the angular velocity route. The former reflects the fact that the integrated diffusive mode dominates the phase at low frequency, where the phase likelihood has its leverage, so that the singular common mode is very visible in phase; the latter reflects the same low-frequency emphasis working the other way, since the quadratic timing projection removes part of what would distinguish a stationary from a slowly diffusing process, whereas the angular velocity route retains more of the exchange-mode dynamics on which the minimal model is tested.

Figure~\ref{fig:phase_velocity_posteriors} compares the crust-only posteriors obtained along the two routes when each realization is analyzed with its generating model. For the singular realization the two posteriors overlap closely. $Q_{\rm c}$ is constrained to the same values along both routes, and $\tau$, $r$ and $Q_{\rm s}$ are prior-dominated along both. For the minimal realization the two posteriors differ markedly. Both lie on the ridge $A_{\rm min}\propto\tau$ discussed in Section~\ref{sec:minimal_simulations}, but the angular velocity posterior is displaced along the ridge toward longer relaxation times and larger amplitudes, by close to an order of magnitude in both coordinates relative to the phase posterior, while the ratio $A_{\rm min}/\tau$ that sets the short-time diffusion of the lag is nearly unchanged. It is worth noting that this displacement is not primarily a consequence of the window length. The windows of 20 days are short compared with the relaxation time, and the covariance of the slopes differs from the instantaneous covariance by only a few percent. Rather, the direction of the shift is the one in which the data are least informative, and the angular velocity route is less informative there than the phase route for three compounding reasons. First, the cadence compression from 101 phases to 20 slopes removes the short-lag pairs that best resolve the curvature of the OU covariance near zero lag, which is what separates $\tau$ from $A_{\rm min}/\tau$. Second, the window slopes have a lower signal-to-noise ratio per point than the phases from which they are formed. Third, an ideal control in which the instantaneous angular velocity is observed at the same twenty epochs with the same nominal errors recovers the generating parameters, which locates the loss in the averaging and cadence compression of $\boldsymbol H$ rather than in the sample count or the noise level alone. The singular posterior is less affected because its best-constrained coordinate, $Q_{\rm c}$, sets the amplitude of the dominant diffusive mode, which is a low-frequency feature that both routes retain, whereas the minimal posterior is constrained through the exchange mode, whose time scale must be resolved against the sampling. The shift is seen in a single realization and the two posteriors are compatible within their widths along the ridge; we do not read it as a bias of one route relative to the other, but as an indication that, for a stationary process, the phase route retains more of the information that pins the relaxation time.

The measurement equation can also carry systematic structure that is not part of either model. A phase step, an epoch offset or an unmodelled low-order trend is transformed differently along the two routes. The timing projection absorbs slow trends, whereas differentiation suppresses them but converts a phase step into a localized, high-amplitude feature in the derived angular velocities. Such effects are best interpreted through the actual linear operator and its propagated covariance rather than through a generic description of differentiation as a high-pass filter.

Radio timing data are phases, and the physical question is posed in angular velocity. In summary, we find that both routes/domains can both lead to the same qualitative conclusions about what full-state and crust-only observations can establish. The phase route analyses the data as they are delivered, with a likelihood that accounts exactly for the timing model and for the integration-induced nonstationarity, and it retains the short-lag information that a window-averaged angular velocity series compresses away. The angular velocity route places the analysis in the variables in which the physics is stated, at the cost of an intermediate processing step whose averaging and cadence compression must be kept in mind when interpreting the posteriors. Neither route recovers the information lost with a hidden superfluid state.

\section{What pulsar timing can do?}
\label{sec:testing_conservation}

The question that motivates this work is whether radio pulsar timing can tell us whether the spin noise of a neutron star is stationary, and thereby whether the angular momentum of the star is conserved internally or exchanged with the outside through independent stochastic torques on its components. The preceding sections allow a conservative answer to be given, in three parts, on what timing can do on its own, on what it cannot, and on what becomes possible with external information.

Radio timing on its own can be analyzed exactly for this purpose. The likelihood of Section~\ref{sec:likelihood} is written for the phases that timing delivers
and it scales linearly with the number of observations. The simulations show that this likelihood reconstructs the crust phase and constrains the parameters that set the amplitude of the crust spin noise, the crust diffusion coefficient in the singular model and the torque strength $A_{\rm min}/\tau$ in the minimal model, from the crust alone. Timing also responds to the physics we are after. When the generating process has a diffusive total angular momentum, the crust phase carries an imprint of it that the singular model picks up, and the sign of the Bayes factor points to the singular model in most realizations. This is what one would hope for from a method that is to be applied to real pulsars, and it holds regardless of whether the phases are analyzed directly or first differentiated into spin frequencies.

Radio timing on its own cannot, for observations of the span and quality simulated here, establish either alternative with confidence. The crust-only Bayes factors are of order unity for a singular generating process, corresponding to odds of a few to one, and consistent with zero for a minimal generating process, and their magnitudes shift with the prior volume assigned to the singular model. The reason is that both models describe the observed crust phase equally well over a substantial part of their parameter space. The crust phase of the minimal model is fixed by two combinations of its parameters, and the singular model reproduces it whenever its diffusive mode is weak enough to hide below the measurement noise. The parameters that distinguish the two pictures physically, the partition of the stochastic torques between the components in the singular model and the superfluid inertia fraction in the minimal one, do not enter the crust phase covariance and are returned as their priors. The crust-only comparison therefore reduces to a question of which model has the smaller effective prior volume near the observed crust covariance, which is a statement about the priors as much as about the star. A crust-only preference of the magnitude found here should accordingly be reported together with its prior dependence and not be read as a detection of a diffusive mode or of angular momentum conservation.

External information changes the problem, and this is where the minimal model becomes falsifiable. Because its crust and superfluid phases are tied by the conservation law \eqref{eq:minimal_eye}, the minimal model predicts the superfluid phase from the crust phase up to the single factor $-x_{\rm c}/x_{\rm s}$ (Section~\ref{subsec:conditioning}). An independent estimate of the superfluid inertia fraction, such as that inferred from the angular momentum reservoir required by glitches in the same pulsar \citep{2012PhRvL.109x1103A, Haskell:2015jra, 2020A&A...642A.223M}, fixes this factor and turns the minimal model into a definite prediction for a quantity that timing does not observe, with a well-defined uncertainty inherited from the crust reconstruction. Any independent measurement of the superfluid rotation, however sparse, then tests the prediction directly, and continuous gravitational waves from the interior of the star are the natural candidate for such a measurement \citep{Jaranowski:1998qm,2021arXiv210403137H,Ming:2021xtz, 2021arXiv210403137H, 2022Univ....8..619H, Dergachev:2025hwp, Antonelli:2025vhw}. The singular model makes no comparably rigid prediction, since its superfluid phase depends on the partition of the stochastic torques, which the crust does not constrain. Nonetheless, an independent measurement of the superfluid moment of inertia fraction or the superfluid phase at a few epochs can be used to constrain the parameter space. Whether a glitch-derived inertia fraction refers to the same effective superfluid component that participates in spin wandering is a question of neutron-star physics that the present framework does not settle, and we treat such an identification as a hypothesis to be tested rather than as an input.

Two further directions follow from these considerations. A pulsar with both a long timing baseline and a glitch-derived inertia fraction is the immediate application.
Beyond individual pulsars, the stationarity test could be carried out across a population. Isolated pulsars are the natural sample, since a binary companion provides an external channel through which angular momentum may be exchanged stochastically with the pulsar, and a preference for the singular model in isolated pulsars would be harder to attribute to external torques. Millisecond pulsars whose timing noise is loud compared with the gravitational wave background are of particular interest, since they combine long, dense baselines with a spin noise that can be studied on its own. Such a study would ask whether the crust-only Bayes factors of a population lean consistently in one direction.

\section{Conclusions}
\label{sec:conclusions}

We have developed a scalable Gaussian process method for pulsar spin noise that works directly with the phases, or timing residuals, delivered by radio pulsar timing. The method is built on the exact analytical solutions of the singular, nonstationary and the minimal, stationary two-component models of a neutron star \citep{Bernardo:2026opv}.
The crust phase covariances of the two models turn out to be semiseparable, of rank three and two respectively, which allows the likelihood to be evaluated with a cost that grows linearly with the number of observations, and the deterministic timing model is marginalized analytically within the same framework. The result is an exact likelihood for the observable of radio timing that is anchored to the dynamics of the two components.

The physical question
is whether the spin noise of a neutron star is stationary, which within two-component models is the question of whether the total angular momentum of the star is conserved internally. It is posed in the angular velocities, or spin frequencies, whereas timing observes the crust phase. We have therefore had to deal with several operations that stand between the physics and the data. Integration turns a stationary angular velocity into a nonstationary phase, so that a wandering phase is not by itself evidence for a random walk in the spin \citep{Antonelli:2022gqw, Bernardo:2026opv}. The least-squares removal of the timing model suppresses the lowest frequencies and is itself a source of nonstationarity in the residuals \citep{Lee:2012bf, vanHaasteren:2012hj, Allen:2025waa}. Differentiation of the phases into local spin frequencies averages the state over finite windows and thins the sampling. We have treated each of these operations exactly, and we have found that none of them changes the qualitative answer to the physical question. The stationarity or nonstationarity of the spin noise is neither created nor destroyed on the way from the state to the data. What the operations change is how much of the distinction survives in a finite, noisy data set, and here the phase route retains the short-lag information that a window-averaged spin-frequency series compresses away, which matters most for the stationary model, whose relaxation time and amplitude are otherwise degenerate over observation spans of a few relaxation times.

The simulations give the same central result as their angular velocity counterparts.
When both the crust and the superfluid are observed, the parameters of the generating process are recovered, both phases are reconstructed, and the stationary and nonstationary models are told apart decisively. When only the crust is observed, as in radio timing, the crust phase is reconstructed and the amplitude of its spin noise is measured, but the two models describe the observed crust equally well over much of their parameter space. A singular generating process leaves a weak preference in its favour, with posterior odds of a few to one that shift with the prior volume of the singular model. A minimal generating process leaves no preference. Radio timing alone, for observations of the span and quality simulated here, does not establish either alternative with confidence, and the crust-only Bayes factors should be read as prior-limited.

The minimal model nevertheless makes a prediction that timing alone does not. Because its stochastic torques conserve angular momentum, it ties the superfluid phase to the crust phase up to a single factor set by the superfluid inertia fraction. An external estimate of this fraction, as inferred from glitches in the same pulsar, turns the minimal model into a falsifiable prediction for the hidden rotational state, which a continuous gravitational wave observation of the interior could test.





\appendix

\section{Kernel derivation and orientation}
\label{app:kernels}

For deterministic initial modes the singular phase fluctuations can be written
\begin{align}
 \delta\phi_{\rm c}(t)&=\frac{\tau}{T_\Sigma I}\int_{t_0}^t\{(t-\lambda)\xi_+(\lambda)+\tau_{\rm s}[1-\e^{-(t-\lambda)/\tau}]\xi_-(\lambda)\}\dd\lambda,\\
 \delta\phi_{\rm s}(t)&=\frac{\tau}{T_\Sigma I}\int_{t_0}^t\{(t-\lambda)\xi_+(\lambda)-\tau_{\rm c}[1-\e^{-(t-\lambda)/\tau}]\xi_-(\lambda)\}\dd\lambda.
\end{align}
Contracting the white-noise increments gives (\ref{eq:I1}-\ref{eq:Ksc}). The first index of $\mathcal I_{+-}$ identifies the response evaluated at the first covariance time.  This convention immediately yields the multi-output symmetry $K_{ij}(t,t')=K_{ji}(t',t)$.

\section{Semiseparable factors and solves}
\label{app:ss}

For $i>j$, every polynomial--exponential term in the singular $K_{\rm cc}$ separates into a product of a function of $t_i$ and a function of $t_j$.  Collecting the independent functions gives three columns in $\boldsymbol U$ and $\boldsymbol V$.  The minimal kernel cancels the diffusive polynomial column and requires two.  Exponential centring at the midpoint of the elapsed-time range leaves $\boldsymbol U_i\boldsymbol V_j^{\mathsf T}$ invariant while avoiding overflow.

Explicitly, with $u_i=t_i-t_0$ and $G_\phi=\tau^2/T_\Sigma^2$, the singular crust kernel of equation~\eqref{eq:Kcc} has, for $i>j$,
\begin{equation}
 \boldsymbol U_i=\begin{pmatrix}\e^{-u_i/\tau}&u_i&1\end{pmatrix},
\end{equation}
and
\begin{align}
 V_j^{(1)}&=G_\phi\Big\{Q_\times\tau_{\rm s}\left[\tau(u_j+\tau)-\tau^2\e^{u_j/\tau}\right] \nonumber\\
 &\qquad\quad+Q_-\tau_{\rm s}^2\left[\tau-\tfrac{\tau}{2}\left(\e^{u_j/\tau}+\e^{-u_j/\tau}\right)\right]\Big\},\\
 V_j^{(2)}&=G_\phi\Big\{\tfrac{Q_+}{2}u_j^2+Q_\times\tau_{\rm s}\left[u_j-\tau+\tau\e^{-u_j/\tau}\right]\Big\},\\
 V_j^{(3)}&=G_\phi\Big\{-\tfrac{Q_+}{6}u_j^3+Q_\times\tau_{\rm s}\left[\tau u_j-\tau^2+\tau^2\e^{-u_j/\tau}\right] \nonumber\\
 &\qquad\quad+Q_-\tau_{\rm s}^2\left[u_j-\tau+\tau\e^{-u_j/\tau}\right]\Big\},
\end{align}
which follow from grouping the terms of $\mathcal I_1$, $\mathcal I_2$ and $\mathcal I_3$ according to their dependence on the later time. The diagonal is
\begin{align}
 a_i&=\sigma_i^2+G_\phi\Big[Q_+\tfrac{u_i^3}{3}+Q_\times\tau_{\rm s}\left(u_i^2-2\tau^2+2\tau(u_i+\tau)\e^{-u_i/\tau}\right) \nonumber\\
 &\qquad\quad+Q_-\tau_{\rm s}^2\left(u_i-\tfrac{3\tau}{2}+2\tau\e^{-u_i/\tau}-\tfrac{\tau}{2}\e^{-2u_i/\tau}\right)\Big].
\end{align}
For the minimal crust kernel of equation~\eqref{eq:Amin},
\begin{align}
 \boldsymbol U_i&=\begin{pmatrix}\e^{-u_i/\tau}&1\end{pmatrix},\\
 V_j^{(1)}&=\frac{A_{\rm min}}{2}\tau^2\left(1-\e^{u_j/\tau}\right),\\
 V_j^{(2)}&=\frac{A_{\rm min}}{2}\left(2\tau u_j-\tau^2+\tau^2\e^{-u_j/\tau}\right),
\end{align}
and $a_i=\sigma_i^2+\tfrac{A_{\rm min}}{2}\mathcal H(u_i,u_i)$. The growing exponentials $\e^{u_j/\tau}$ in $\boldsymbol V_j$ are paired with the decaying $\e^{-u_i/\tau}$ in $\boldsymbol U_i$; the centring mentioned above rescales both by a common factor so that neither overflows.

For a right-hand side $b_i$, the forward solve uses
\begin{equation}
 z_i=\frac{b_i-\boldsymbol U_i\boldsymbol f_i}{\sqrt{d_i}},\qquad
 \boldsymbol f_{i+1}=\boldsymbol f_i+\boldsymbol W_i^{\mathsf T}\sqrt{d_i}z_i,
\end{equation}
and the backward solve uses
\begin{equation}
 x_i=\frac{z_i}{\sqrt{d_i}}-\boldsymbol W_i\boldsymbol g_i,\qquad
 \boldsymbol g_{i-1}=\boldsymbol g_i+\boldsymbol U_i^{\mathsf T}x_i.
\end{equation}
The same factorization solves simultaneously for $\by$ and the three columns of $\bX$.

\section{Analytic and numerical verification}
\label{app:verification}

The implementation was checked at three levels.
Differentiation verifies
\begin{equation}
 \frac{\partial^2K_{ij}(t,t')}{\partial t\,\partial t'}=\Cov[\Omega_i(t),\Omega_j(t')]
\end{equation}
piecewise away from the covariance cusp.  Centred finite differences reproduce the same angular velocity covariances numerically.  Finally, exact-covariance Monte Carlo phase simulations were differentiated and compared with direct angular velocity simulations; sample covariances converge to the analytic blocks.  Dense and semiseparable phase likelihoods agree to numerical precision, covariance matrices are positive semidefinite, and oriented cross-blocks satisfy the required transpose relation.

\section{Evidence settings and reproducibility}
\label{app:evidence}

Full-state ensembles use bounded Laplace evidence from the exact dense multi-output phase Gaussian process.  The radio-only headline uses stepping-stone evidence with normalized bounded priors.  The cached high-statistics products contain 1000 full-state realizations, 500 stepping-stone radio realizations and 200 singular-prior sensitivity realizations for each generator and domain.  Their metadata record deterministic seeds, bounds, configuration fingerprints, software context and acceptance diagnostics.  The high-statistics production ladder used 80 retained steps, 40 burn-in steps, thinning by four and 1024 prior samples; smaller real-data applications may use longer chains.  Laplace radio evidences are retained only as local-geometry diagnostics.


\begin{thebibliography}{}
\makeatletter
\relax
\def\mn@urlcharsother{\let\do\@makeother \do\$\do\&\do\#\do\^\do\_\do\%\do\~}
\def\mn@doi{\begingroup\mn@urlcharsother \@ifnextchar [ {\mn@doi@}
  {\mn@doi@[]}}
\def\mn@doi@[#1]#2{\def\@tempa{#1}\ifx\@tempa\@empty \href
  {http://dx.doi.org/#2} {doi:#2}\else \href {http://dx.doi.org/#2} {#1}\fi
  \endgroup}
\def\mn@eprint#1#2{\mn@eprint@#1:#2::\@nil}
\def\mn@eprint@arXiv#1{\href {http://arxiv.org/abs/#1} {{\tt arXiv:#1}}}
\def\mn@eprint@dblp#1{\href {http://dblp.uni-trier.de/rec/bibtex/#1.xml}
  {dblp:#1}}
\def\mn@eprint@#1:#2:#3:#4\@nil{\def\@tempa {#1}\def\@tempb {#2}\def\@tempc
  {#3}\ifx \@tempc \@empty \let \@tempc \@tempb \let \@tempb \@tempa \fi \ifx
  \@tempb \@empty \def\@tempb {arXiv}\fi \@ifundefined
  {mn@eprint@\@tempb}{\@tempb:\@tempc}{\expandafter \expandafter \csname
  mn@eprint@\@tempb\endcsname \expandafter{\@tempc}}}

\bibitem[\protect\citeauthoryear{Agazie et~al.}{Agazie
  et~al.}{2023}]{NANOGrav:2023gor}
Agazie G.,  et~al., 2023, \mn@doi [Astrophys. J. Lett.]
  {10.3847/2041-8213/acdac6}, 951, L8

\bibitem[\protect\citeauthoryear{Allen, von Blanckenburg  \& Olum}{Allen
  et~al.}{2026}]{Allen:2025waa}
Allen B.,  von Blanckenburg A.~L.,   Olum K.~D.,  2026, \mn@doi [Phys. Rev. D]
  {10.1103/2dft-4rjj}, 113, 102001

\bibitem[\protect\citeauthoryear{{Ambikasaran}, {Foreman-Mackey}, {Greengard},
  {Hogg}  \& {O'Neil}}{{Ambikasaran} et~al.}{2015}]{2015ITPAM..38..252A}
{Ambikasaran} S.,  {Foreman-Mackey} D.,  {Greengard} L.,  {Hogg} D.~W.,
  {O'Neil} M.,  2015, \mn@doi [IEEE Transactions on Pattern Analysis and
  Machine Intelligence] {10.1109/TPAMI.2015.2448083}, \href
  {https://ui.adsabs.harvard.edu/abs/2015ITPAM..38..252A} {38, 252}

\bibitem[\protect\citeauthoryear{{Andersson}, {Glampedakis}, {Ho}  \&
  {Espinoza}}{{Andersson} et~al.}{2012}]{2012PhRvL.109x1103A}
{Andersson} N.,  {Glampedakis} K.,  {Ho} W.~C.~G.,   {Espinoza} C.~M.,  2012,
  \mn@doi [\prl] {10.1103/PhysRevLett.109.241103}, \href
  {https://ui.adsabs.harvard.edu/abs/2012PhRvL.109x1103A} {109, 241103}

\bibitem[\protect\citeauthoryear{Antonelli, Basu  \& Haskell}{Antonelli
  et~al.}{2023}]{Antonelli:2022gqw}
Antonelli M.,  Basu A.,   Haskell B.,  2023, \mn@doi [Mon. Not. Roy. Astron.
  Soc.] {10.1093/mnras/stad256}, 520, 2813

\bibitem[\protect\citeauthoryear{Antonelli, Basu  \& Haskell}{Antonelli
  et~al.}{2025}]{Antonelli:2025vhw}
Antonelli M.,  Basu A.,   Haskell B.,  2025, \mn@doi [Publ. Astron. Soc.
  Austral.] {10.1017/pasa.2025.10086}, 42, e118

\bibitem[\protect\citeauthoryear{Antoniadis et~al.}{Antoniadis
  et~al.}{2023}]{EPTA:2023fyk}
Antoniadis J.,  et~al., 2023, \mn@doi [Astron. Astrophys.]
  {10.1051/0004-6361/202346844}, 678, A50

\bibitem[\protect\citeauthoryear{{Baym}, {Pethick}, {Pines}  \&
  {Ruderman}}{{Baym} et~al.}{1969}]{1969Natur.224..872B}
{Baym} G.,  {Pethick} C.,  {Pines} D.,   {Ruderman} M.,  1969, \mn@doi [Nature]
  {10.1038/224872a0}, 224, 872

\bibitem[\protect\citeauthoryear{Bernardo}{Bernardo}{2026}]{Bernardo:2026opv}
Bernardo R.~C.,  2026, {Stochastic problems in pulsar timing} (\mn@eprint
  {arXiv} {2604.08373})

\bibitem[\protect\citeauthoryear{{Chamel}}{{Chamel}}{2013}]{2013PhRvL.110a1101C}
{Chamel} N.,  2013, \mn@doi [\prl] {10.1103/PhysRevLett.110.011101}, \href
  {https://ui.adsabs.harvard.edu/abs/2013PhRvL.110a1101C} {110, 011101}

\bibitem[\protect\citeauthoryear{Dergachev \& Papa}{Dergachev \&
  Papa}{2025}]{Dergachev:2025hwp}
Dergachev V.,  Papa M.~A.,  2025, \mn@doi [Phys. Rev. D] {10.1103/psrl-y44w},
  112, 042005

\bibitem[\protect\citeauthoryear{Dong, Melatos, O'Neill, Meyers  \& Boek}{Dong
  et~al.}{2026}]{Dong:2025nho}
Dong W.,  Melatos A.,  O'Neill N.~J.,  Meyers P.~M.,   Boek D.~K.,  2026,
  \mn@doi [Mon. Not. Roy. Astron. Soc.] {10.1093/mnras/staf2053}, 545, staf2053

\bibitem[\protect\citeauthoryear{{Dunn} et~al.,}{{Dunn}
  et~al.}{2025}]{2025MNRAS.541.1792D}
{Dunn} L.,  et~al., 2025, \mn@doi [\mnras] {10.1093/mnras/staf1064}, \href
  {https://ui.adsabs.harvard.edu/abs/2025MNRAS.541.1792D} {541, 1792}

\bibitem[\protect\citeauthoryear{Edwards, Hobbs  \& Manchester}{Edwards
  et~al.}{2006}]{Edwards:2006zg}
Edwards R.~T.,  Hobbs G.~B.,   Manchester R.~N.,  2006, \mn@doi [Mon. Not. Roy.
  Astron. Soc.] {10.1111/j.1365-2966.2006.10870.x}, 372, 1549

\bibitem[\protect\citeauthoryear{Foreman-Mackey, Hogg, Lang  \&
  Goodman}{Foreman-Mackey et~al.}{2013}]{emcee2013}
Foreman-Mackey D.,  Hogg D.~W.,  Lang D.,   Goodman J.,  2013, \mn@doi [Publ.
  Astron. Soc. Pac.] {10.1086/670067}, 125, 306

\bibitem[\protect\citeauthoryear{{Foreman-Mackey}, {Agol}, {Ambikasaran}  \&
  {Angus}}{{Foreman-Mackey} et~al.}{2017}]{2017AJ....154..220F}
{Foreman-Mackey} D.,  {Agol} E.,  {Ambikasaran} S.,   {Angus} R.,  2017,
  \mn@doi [Astron. J.] {10.3847/1538-3881/aa9332}, \href
  {https://ui.adsabs.harvard.edu/abs/2017AJ....154..220F} {154, 220}

\bibitem[\protect\citeauthoryear{Goncharov, Zhu  \& Thrane}{Goncharov
  et~al.}{2020}]{Goncharov:2019mei}
Goncharov B.,  Zhu X.-J.,   Thrane E.,  2020, \mn@doi [Mon. Not. Roy. Astron.
  Soc.] {10.1093/mnras/staa2081}, 497, 3264

\bibitem[\protect\citeauthoryear{{Groth}}{{Groth}}{1975}]{1975ApJS...29..453G}
{Groth} E.~J.,  1975, \mn@doi [\apjs] {10.1086/190354}, \href
  {https://ui.adsabs.harvard.edu/abs/1975ApJS...29..453G} {29, 453}

\bibitem[\protect\citeauthoryear{Haskell \& Melatos}{Haskell \&
  Melatos}{2015}]{Haskell:2015jra}
Haskell B.,  Melatos A.,  2015, \mn@doi [Int. J. Mod. Phys. D]
  {10.1142/S0218271815300086}, 24, 1530008

\bibitem[\protect\citeauthoryear{{Haskell} \& {Schwenzer}}{{Haskell} \&
  {Schwenzer}}{2021}]{2021arXiv210403137H}
{Haskell} B.,  {Schwenzer} K.,  2021, \mn@doi [arXiv e-prints]
  {10.48550/arXiv.2104.03137}, \href
  {https://ui.adsabs.harvard.edu/abs/2021arXiv210403137H} {p. arXiv:2104.03137}

\bibitem[\protect\citeauthoryear{{Haskell}, {Antonelli}  \&
  {Pizzochero}}{{Haskell} et~al.}{2022}]{2022Univ....8..619H}
{Haskell} B.,  {Antonelli} M.,   {Pizzochero} P.,  2022, \mn@doi [Universe]
  {10.3390/universe8120619}, \href
  {https://ui.adsabs.harvard.edu/abs/2022Univ....8..619H} {8, 619}

\bibitem[\protect\citeauthoryear{Hazboun, Romano  \& Smith}{Hazboun
  et~al.}{2019}]{Hazboun:2019vhv}
Hazboun J.~S.,  Romano J.~D.,   Smith T.~L.,  2019, \mn@doi [Phys. Rev. D]
  {10.1103/PhysRevD.100.104028}, 100, 104028

\bibitem[\protect\citeauthoryear{Hewish, Bell, Pilkington, Scott  \&
  Collins}{Hewish et~al.}{1968}]{Hewish:1968bj}
Hewish A.,  Bell S.~J.,  Pilkington J. D.~H.,  Scott P.~F.,   Collins R.~A.,
  1968, \mn@doi [Nature] {10.1038/217709a0}, 217, 709

\bibitem[\protect\citeauthoryear{Hobbs, Edwards  \& Manchester}{Hobbs
  et~al.}{2006}]{Hobbs:2006cd}
Hobbs G.,  Edwards R.,   Manchester R.,  2006, \mn@doi [Mon. Not. Roy. Astron.
  Soc.] {10.1111/j.1365-2966.2006.10302.x}, 369, 655

\bibitem[\protect\citeauthoryear{{Hobbs}, {Lyne}  \& {Kramer}}{{Hobbs}
  et~al.}{2010}]{2010MNRAS.402.1027H}
{Hobbs} G.,  {Lyne} A.~G.,   {Kramer} M.,  2010, \mn@doi [\mnras]
  {10.1111/j.1365-2966.2009.15938.x}, \href
  {https://ui.adsabs.harvard.edu/abs/2010MNRAS.402.1027H} {402, 1027}

\bibitem[\protect\citeauthoryear{Hulse \& Taylor}{Hulse \&
  Taylor}{1975}]{Hulse:1974eb}
Hulse R.~A.,  Taylor J.~H.,  1975, \mn@doi [Astrophys. J. Lett.]
  {10.1086/181708}, 195, L51

\bibitem[\protect\citeauthoryear{{Jankowski} et~al.,}{{Jankowski}
  et~al.}{2019}]{2019MNRAS.484.3691J}
{Jankowski} F.,  et~al., 2019, \mn@doi [\mnras] {10.1093/mnras/sty3390}, \href
  {https://ui.adsabs.harvard.edu/abs/2019MNRAS.484.3691J} {484, 3691}

\bibitem[\protect\citeauthoryear{Jaranowski, Krolak  \& Schutz}{Jaranowski
  et~al.}{1998}]{Jaranowski:1998qm}
Jaranowski P.,  Krolak A.,   Schutz B.~F.,  1998, \mn@doi [Phys. Rev. D]
  {10.1103/PhysRevD.58.063001}, 58, 063001

\bibitem[\protect\citeauthoryear{{Kimpson}, {O'Neill}, {Meyers}  \&
  {Melatos}}{{Kimpson} et~al.}{2025}]{2025arXiv251011077K}
{Kimpson} T.,  {O'Neill} N.~J.,  {Meyers} P.~M.,   {Melatos} A.,  2025, \mn@doi
  [arXiv e-prints] {10.48550/arXiv.2510.11077}, \href
  {https://ui.adsabs.harvard.edu/abs/2025arXiv251011077K} {p. arXiv:2510.11077}

\bibitem[\protect\citeauthoryear{Kramer et~al.}{Kramer
  et~al.}{2021}]{Kramer:2021jcw}
Kramer M.,  et~al., 2021, \mn@doi [Phys. Rev. X] {10.1103/PhysRevX.11.041050},
  11, 041050

\bibitem[\protect\citeauthoryear{{Lasky}, {Melatos}, {Ravi}  \&
  {Hobbs}}{{Lasky} et~al.}{2015}]{2015MNRAS.449.3293L}
{Lasky} P.~D.,  {Melatos} A.,  {Ravi} V.,   {Hobbs} G.,  2015, \mn@doi [\mnras]
  {10.1093/mnras/stv540}, \href
  {https://ui.adsabs.harvard.edu/abs/2015MNRAS.449.3293L} {449, 3293}

\bibitem[\protect\citeauthoryear{Lee, Bassa, Karuppusamy, Kramer, Smits  \&
  Stappers}{Lee et~al.}{2012}]{Lee:2012bf}
Lee K.~J.,  Bassa C.~G.,  Karuppusamy R.,  Kramer M.,  Smits R.,   Stappers
  B.~W.,  2012, \mn@doi [Mon. Not. Roy. Astron. Soc.]
  {10.1111/j.1365-2966.2012.21070.x}, 423, 2642

\bibitem[\protect\citeauthoryear{Lorimer}{Lorimer}{2001}]{Lorimer:2001vd}
Lorimer D.~R.,  2001, \mn@doi [Living Rev. Rel.] {10.12942/lrr-2001-5}, 4, 5

\bibitem[\protect\citeauthoryear{{Lower} et~al.,}{{Lower}
  et~al.}{2020}]{2020MNRAS.494..228L}
{Lower} M.~E.,  et~al., 2020, \mn@doi [\mnras] {10.1093/mnras/staa615}, \href
  {https://ui.adsabs.harvard.edu/abs/2020MNRAS.494..228L} {494, 228}

\bibitem[\protect\citeauthoryear{{Mackay}}{{Mackay}}{2003}]{2003itil.book.....M}
{Mackay} D. J.~C.,  2003, {Information Theory, Inference and Learning
  Algorithms}

\bibitem[\protect\citeauthoryear{Manchester}{Manchester}{2015}]{Manchester:2015mda}
Manchester R.~N.,  2015, \mn@doi [Int. J. Mod. Phys. D]
  {10.1142/S0218271815300189}, 24, 1530018

\bibitem[\protect\citeauthoryear{Melatos \& Link}{Melatos \&
  Link}{2014}]{Melatos:2013rca}
Melatos A.,  Link B.,  2014, \mn@doi [Mon. Not. Roy. Astron. Soc.]
  {10.1093/mnras/stt1828}, 437, 21

\bibitem[\protect\citeauthoryear{Meyers, Melatos  \& O'Neill}{Meyers
  et~al.}{2021a}]{Meyers:2021myb}
Meyers P.~M.,  Melatos A.,   O'Neill N.~J.,  2021a, \mn@doi [Mon. Not. Roy.
  Astron. Soc.] {10.1093/mnras/stab262}, 502, 3113

\bibitem[\protect\citeauthoryear{Meyers, O'Neill, Melatos  \& Evans}{Meyers
  et~al.}{2021b}]{Meyers:2021slh}
Meyers P.~M.,  O'Neill N.~J.,  Melatos A.,   Evans R.~J.,  2021b, \mn@doi [Mon.
  Not. Roy. Astron. Soc.] {10.1093/mnras/stab1952}, 506, 3349

\bibitem[\protect\citeauthoryear{Miles et~al.}{Miles
  et~al.}{2025}]{Miles:2024seg}
Miles M.~T.,  et~al., 2025, \mn@doi [Mon. Not. Roy. Astron. Soc.]
  {10.1093/mnras/stae2571}, 536, 1489

\bibitem[\protect\citeauthoryear{Ming, Papa, Eggenstein, Machenschalk,
  Steltner, Prix, Allen  \& Behnke}{Ming et~al.}{2022}]{Ming:2021xtz}
Ming J.,  Papa M.~A.,  Eggenstein H.-B.,  Machenschalk B.,  Steltner B.,  Prix
  R.,  Allen B.,   Behnke O.,  2022, \mn@doi [Astrophys. J.]
  {10.3847/1538-4357/ac35cb}, 925, 8

\bibitem[\protect\citeauthoryear{{Montoli}, {Antonelli}, {Magistrelli}  \&
  {Pizzochero}}{{Montoli} et~al.}{2020}]{2020A&A...642A.223M}
{Montoli} A.,  {Antonelli} M.,  {Magistrelli} F.,   {Pizzochero} P.~M.,  2020,
  \mn@doi [\aap] {10.1051/0004-6361/202038340}, \href
  {https://ui.adsabs.harvard.edu/abs/2020A&A...642A.223M} {642, A223}

\bibitem[\protect\citeauthoryear{O'Neill, Meyers  \& Melatos}{O'Neill
  et~al.}{2024}]{ONeill:2024uiw}
O'Neill N.~J.,  Meyers P.~M.,   Melatos A.,  2024, \mn@doi [Mon. Not. Roy.
  Astron. Soc.] {10.1093/mnras/stae770}, 530, 4648

\bibitem[\protect\citeauthoryear{Pilkington, Hewish, Bell  \& Cole}{Pilkington
  et~al.}{1968}]{Pilkington:1968bk}
Pilkington J. D.~H.,  Hewish A.,  Bell S.~J.,   Cole T.~W.,  1968, \mn@doi
  [Nature] {10.1038/218126a0}, 218, 126

\bibitem[\protect\citeauthoryear{Pitrou \& Cusin}{Pitrou \&
  Cusin}{2025}]{Pitrou:2024scp}
Pitrou C.,  Cusin G.,  2025, \mn@doi [Phys. Rev. D]
  {10.1103/PhysRevD.111.083544}, 111, 083544

\bibitem[\protect\citeauthoryear{Reardon et~al.}{Reardon
  et~al.}{2023}]{Reardon:2023gzh}
Reardon D.~J.,  et~al., 2023, \mn@doi [Astrophys. J. Lett.]
  {10.3847/2041-8213/acdd02}, 951, L6

\bibitem[\protect\citeauthoryear{{Shannon} \& {Cordes}}{{Shannon} \&
  {Cordes}}{2010}]{2010ApJ...725.1607S}
{Shannon} R.~M.,  {Cordes} J.~M.,  2010, \mn@doi [\apj]
  {10.1088/0004-637X/725/2/1607}, \href
  {https://ui.adsabs.harvard.edu/abs/2010ApJ...725.1607S} {725, 1607}

\bibitem[\protect\citeauthoryear{Trotta}{Trotta}{2008}]{Trotta:2008qt}
Trotta R.,  2008, \mn@doi [Contemp. Phys.] {10.1080/00107510802066753}, 49, 71

\bibitem[\protect\citeauthoryear{Verbiest, Vigeland, Porayko, Chen  \&
  Reardon}{Verbiest et~al.}{2024}]{Verbiest:2024nid}
Verbiest J. P.~W.,  Vigeland S.~J.,  Porayko N.~K.,  Chen S.,   Reardon D.~J.,
  2024, \mn@doi [Results Phys.] {10.1016/j.rinp.2024.107719}, 61, 107719

\bibitem[\protect\citeauthoryear{Xie, Lewis, Fan, Kuo  \& Chen}{Xie
  et~al.}{2011}]{Xie2011SteppingStone}
Xie W.,  Lewis P.~O.,  Fan Y.,  Kuo L.,   Chen M.-H.,  2011, \mn@doi
  [Systematic Biology] {10.1093/sysbio/syq085}, 60, 150

\bibitem[\protect\citeauthoryear{Xu et~al.}{Xu et~al.}{2023}]{Xu:2023wog}
Xu H.,  et~al., 2023, \mn@doi [Res. Astron. Astrophys.]
  {10.1088/1674-4527/acdfa5}, 23, 075024

\bibitem[\protect\citeauthoryear{{Yu} \& {Allen}}{{Yu} \&
  {Allen}}{2025}]{2025arXiv251208666Y}
{Yu} W.-W.,  {Allen} B.,  2025, \mn@doi [arXiv e-prints]
  {10.48550/arXiv.2512.08666}, \href
  {https://ui.adsabs.harvard.edu/abs/2025arXiv251208666Y} {p. arXiv:2512.08666}

\bibitem[\protect\citeauthoryear{van Haasteren \& Levin}{van Haasteren \&
  Levin}{2013}]{vanHaasteren:2012hj}
van Haasteren R.,  Levin Y.,  2013, \mn@doi [Mon. Not. Roy. Astron. Soc.]
  {10.1093/mnras/sts097}, 428, 1147

\makeatother
\end{thebibliography}

\label{lastpage}
\end{document}